\documentclass{aa}  

\usepackage[colorlinks=true, linkcolor = blue,
            urlcolor  = blue,
            citecolor = blue,
            anchorcolor = blue, bookmarks=false]{hyperref}
\usepackage{graphicx}
\usepackage{txfonts}
\usepackage{natbib}
\usepackage{enumitem} 
\usepackage{color}
\usepackage{threeparttable}
\usepackage{tikz,amsmath}
\usetikzlibrary{shapes.geometric, arrows}
\usepackage{ulem}
\usepackage{wrapfig}
\usepackage{rotating}
\usepackage{lscape} 
\tikzstyle{process} = [rectangle, minimum width=1.5em, minimum height=3.5em, text centered, draw=blue, fill=gray!10]
\tikzstyle{process2} = [rectangle, minimum width=1.5em, minimum height=3.5em, text centered, draw=white, fill=white]
\tikzstyle{arrow} = [thick,->,>=stealth]
\usepackage{cancel}
\usepackage{ulem}
\bibpunct{(}{)}{;}{f}{}{,} 

\newcommand{\teff}{$T_{\mathrm{eff}}$}

\newcommand{\logg}{\mbox{log \textit{g}}}
 
\newcommand{\vsini}{$v\mathrm{sin}\,i$}
\newcommand{\titan}{\textsc{Titans}}
\newcommand{\spl}{\textit{Splash}}
\newcommand{\Ere}{\textit{Erebus}}

\begin{document}

   \title{Chronology of the chemical enrichment of the old Galactic stellar populations
   \thanks{Chemical abundances and updated ages of the \titan, as well as the data used in Sect.~\ref{sec:discusion}, are available at CDS via anonymous ftp to XXX}
   }


   \author{R. E. Giribaldi\inst{1,2}
         \and
          R. Smiljanic\inst{2}
          }
   \institute{Institut d'Astronomie et d'Astrophysique, Universit\'e libre de Bruxelles, CP 226, Boulebard do Triomphe, 1050 Brussels, Belgium \\
             \email{riano.giribaldi@ubl.be, rianoesc@gmail.com}
   \and
   Nicolaus Copernicus Astronomical Center, Polish Academy of Sciences, ul. Bartycka 18, 00-716, Warsaw, Poland 
             }
             
    \date{Received November, 2022; Accepted 23 December, 2022}

 
  \abstract
   {The Milky Way accreted several smaller satellite galaxies in its history. These mergers added stars and gas to the Galaxy and affected the properties of the pre-existing stellar populations. Stellar chemical abundances and ages are needed to establish the chronological order of events that occur before, during, and after such mergers.}
   {We report precise ages ($\sim$6.5\%) and chemical abundances for the \titan, a sample of old metal-poor dwarfs and subgiants with accurate atmospheric parameters. We also obtain ages with an average precision of 10\% for a selected sample of dwarf stars from the GALAH survey. We used these stars, located within $\sim$1 kiloparsec of the Sun, to analyse the chronology of the chemical evolution of in-situ and accreted metal-poor stellar populations.}
   {We determined ages by isochrone fitting. For the \titan, we determined abundances of Mg, Si, Ca, Ti, Ni, Ba, and Eu using spectrum synthesis. The [Mg/Fe] abundances of the GALAH stars were re-scaled to be consistent with the abundances of the \titan. We separated stellar populations by primarily employing chemical abundances and orbits.}
   {We find that star formation in the so-called Gaia-Enceladus or Gaia-Sausage galaxy, the last major system to merge with the Milky Way, lasted at least 3 billion years and got truncated 9.6 $\pm$ 0.2 billion years ago. This marks with very high precision the last stage of its merging process. We also identified stars of a heated metal-poor in-situ population with virtually null net rotation, probably disturbed by several of the early Milky Way mergers. We show that this population is more metal rich than Gaia-Enceladus at any time.}
   {The sequence of events uncovered in our analysis supports the hypothesis that Gaia-Enceladus truncated the formation of the high-$\alpha$ disc and caused the gas infall that forms the low-$\alpha$ disc, in agreement with theoretical predictions.}

   \keywords{Standards -- Surveys -- Stars: atmospheres -- Stars: fundamental parameters -- Stars: late-type}

   \maketitle
%

\section{Introduction}

According to standard cosmology, large galaxies form hierarchically by merging smaller stellar systems\citep{1984Natur.311..517B,2005Natur.435..629S}. The identification of Milky Way merger remnants saw great progress in recent years thanks to the combination of the astrometric data from the \textit{Gaia} mission \citep{2016A&A...595A...1G} with photometric and spectroscopic data from large stellar surveys such as the Sloan Digital Sky Survey \citep[SDSS,][]{SDSS-DR7}, the Apache Point Observatory Galactic Evolution Experiment \citep[APOGEE,][]{APOGEE}, and the Galactic Archaeology with HERMES survey \citep[GALAH,][]{GALAH}. Such investigations provide evidence of the early assembly of the Milky Way halo from satellite galaxies \citep[e.g.,][]{Naidu2020,an2021ApJ...918...74A,Shank2022} and demonstrated that the Milky Way protodisc formed very early \citep{2020A&A...636A.115D,2020MNRAS.497L...7S,2020MNRAS.492.3241V,2021MNRAS.506.1438K}. Current evidence indicates that the last major merger to have occurred in the Milky Way was with the so-called Gaia-Enceladus or Gaia-Sausage galaxy \citep{helmi2018,Belokurov2018}.

The early mergers between the Milky Way and smaller galaxies entailed not only the accretion of stars but also of gas. Models have shown that this stripped gas has an influence on the star formation history of the accreting galaxy \citep{vincenzo2019MNRAS.487L..47V,belokurov2020MNRAS.494.3880B,Grand2020}. For the Milky Way, it has been shown that a model with two phases of gas infall \citep{chiappini1997ApJ...477..765C,Micali2013} can explain the distinct chemical evolution of the thick and thin discs. Thin disc stars are in general younger and have lower [$\alpha$/Fe]\footnote{This notation gives the abundance by number of a certain element, in a given star, in a logarithmic scale with respect to the Sun: [A/B] = $\log$ [N(A)/N(B)]$_{\star}$ - $\log$ [N(A)/N(B)]$_{\odot}$. For example, [Fe/H] = $-$1.0 means that the ratio of the number of iron atoms to the number of hydrogen atoms is ten times lower in the star than what is observed in the Sun.} ratios compared to thick disc stars \citep{Fuhrmann2011,haywood2013A&A...560A.109H,RecioBlanco2014}. The stellar age-metallicity distribution of the two populations separates at an age of about 8 Gyr \citep{XiangRix22}. 

The delay between the infall that forms the thick disc and the next that forms the thin disc has been estimated to be between 4.5 and 5.5 Gyr \citep{spitoni2019A&A...623A..60S,Spitoni2020}. It has been conjectured that gas from a merger, such as that with Gaia-Enceladus, could have enabled the formation of the low-$\alpha$ thin disc \citep{Brook2007,Bonaca20}, thus being the source of the second infall gas. In this sense, the delay in star formation between the discs can be understood within a gas shock-heating theory \citep{noguchi2018Natur.559..585N}. Quenching of the previous episode of star formation could also be related to the merger of Gaia-Enceladus \citep{vincenzo2019MNRAS.487L..47V}. Furthermore, the Gaia-Enceladus merger also heated up a proto-disc stellar population, creating part of the population that we now call the inner stellar halo \citep{Brook_2003,Bonaca2017,diMatteo2019A&A...632A...4D,belokurov2020MNRAS.494.3880B}. 

The evidence mentioned above indicates that the Gaia-Enceladus merger played a major role in shaping the configuration of the current stellar populations of the Milky Way, although some uncertainty remains in the chronology of all related events \citep[see, however,][for significant improvements in this sense]{Ciuca2021,Ciuca2022}. To determine the relationship between all of these events, it is particularly critical to obtain a precise timing of the last stage of the merger. Previous attempts have obtained estimates between 8 and 10 billion years ago\footnote{Billion years will be replaced by the abbreviation of gigayear, ``Gyr", henceforth.} \citep{helmi2018,gallart2019NatAs...3..932G,montalban2021NatAs...5..640M,belokurov2020MNRAS.494.3880B,Borre22}.

In this work, we precisely date the chemical enrichment of old Galactic stellar populations taking advantage of the \titan\ \citep{giribaldi2021A&A...650A.194G}. This is a sample of 48 relatively nearby dwarf and subgiant stars with metallicities, [Fe/H], between $-3.1$ and $-0.8$~dex. More importantly, we have previously determined values of their effective temperatures (\teff) and surface gravities (\logg) with accuracy below 1\% ($\sim$50 K and $\sim$0.04 dex, in \teff\ and \logg, respectively). This accuracy is unmatched by any other sample of metal-poor stars and is what enables our determination of highly precise ages.
To complement the analysis, we selected a second sample of similarly unevolved stars from the data release 3 (DR3) of the GALactic Archaeology with HERMES (GALAH) survey \citep{buder2021galah}.

Stellar ages were determined by isochrone fitting. Thanks to the accurate parameters, the derived ages have an average precision of $6.5\%$ (900 Myr or 0.9 Gyr) for 45 of the 48 \titan. For the GALAH turnoff stars, the ages have an average precision of $10\%$ ($1$ Gyr). The precision of these ages enables a robust discussion of the chronology of the events in the Galaxy that can be related to the Gaia-Enceladus merger.


This paper is organised as follows. The data, sample, and stellar orbit computations are described in Sect.~\ref{sec:data}. Section~\ref{sec:consistency} describes how the parameter scale of the GALAH sample was brought in agreement with the accurate scale defined by the \titan. The estimation of stellar ages is described in Sect.~\ref{sec:ages}. Section~\ref{sec:abundances} describes the determination of the chemical abundances for the \titan\ and the correction of the Mg abundances of the GALAH sample. Section~\ref{sec:separation} describes the procedure followed to separate stellar populations. Finally, in Sect.~\ref{sec:discusion} we interpret our results and discuss their implications in the context of the Galaxy evolution.

\section{Data, sample, and stellar orbits}
\label{sec:data}

The spectroscopic data used to determine the atmospheric parameters and abundances of the \titan\, are described in \cite{giribaldi2021A&A...650A.194G}. In short, the observations were obtained with the Ultraviolet and Visual Echelle Spectrograph (UVES) \citep{2000SPIE.4008..534D}, at the Very Large Telescope of the European Southern Observatory (ESO), and are publicly available at the ESO data archive\footnote{http://archive.eso.org/wdb/wdb/adp/phase3\_main/form\\}. The spectra include the H$\alpha$ line at 6562.797~\AA, have resolving power R $\geq$ 40~000, and a signal-to-noise ratio (S/N) greater than 100. The sample selection described in \cite{giribaldi2021A&A...650A.194G} resulted in 41 stars, for which the values of \teff\ and \logg\ were estimated with 1\% accuracy. In addition to those stars, in \cite{giribaldi2021A&A...650A.194G} we also provided accurate parameters for seven metal-poor Gaia benchmark stars \citep{Hawkins16,jofre2018RNAAS...2..152J} which were analysed using the same methods. The combination of these two samples is what we discuss in this work and is the sample that we generically refer to as the 48 \titan.

To increase the sample of stars for age determination, additional metal-poor stars were selected from data release three (DR3) of the GALAH survey \citep{buder2021galah}. We first selected stars with [Fe/H] $< -0.5$, $2 <$ \logg\ $< 5$, \teff\ $> 4000$~K, and high quality-atmospheric parameters (cannon\_flag = 0). We excluded stars that are too cool and/or too bright because their analysis is usually more uncertain. More metal-rich stars were excluded to avoid strong contamination of the sample by thin-disc stars. The sample was restricted to stars with Gaia parallax $\varpi \geq 0.2$ mas that have errors smaller than 20\% \citep{2021A&A...649A...2L}, resulting in $12\,405$ stars. We further removed stars with a total\footnote{By total we mean the sum in quadrature of individual velocity components.} Galactic velocity below +210 km s$^{-1}$, to avoid stars with disc kinematics, reducing the sample to 6250 stars. This initial sample was used to train the algorithm that separates stellar populations (Section \ref{sec:separation}). For the main discussion, ages and chemical abundances are used only for dwarf and turnoff stars with \logg\ in the range $3.5 <$ \logg\ $< 4.5$. This is the range where precise ages can be determined. 

To find the optimal way to separate the populations, in the first step of the analysis, we kept in the sample the stars with unreliable abundances of $\alpha$ elements from the GALAH sample (i.e., flag\_alpha\_fe $>$ 0); this is the sample of 6250 stars mentioned above. This choice was motivated by the need to increase the statistics and the density of points in the multidimensional space where the populations are separated. Applying the flag on $\alpha$ abundances reduces the sample to about 1500 stars. In any case, as we discuss below, we verify that the final separation in the chemical diagram of [$\alpha$/Fe] vs.\ [Fe/H] is consistent with the usual findings in the literature. Accreted stars are found to have low-[$\alpha$/Fe] ratios down to a metallicity around [Fe/H] = $-$1.5. For lower metallicities, high- and low-$\alpha$ sequences merge. Furthermore, when we select only stars with the best ages (Section \ref{sec:ages}), the sample remains only with those stars that have reliable abundances, i.e.\ those with flag\_alpha\_fe = 0. This final sample has 208 stars, 25 \titan\ and 183 from GALAH DR3. These stars are presented in Fig.~\ref{fig:kiel}, where the in-situ populations, \spl\ \citep[defined in ][]{belokurov2020MNRAS.494.3880B} and \Ere\footnote{According to the Greek oral poet Hesiod's Theogony, Erebus is the offspring of Chaos.} (defined in this work, see Section \ref{sec:separation}), and the accreted population of Gaia-Enceladus are distinguished. 

For the GALAH sample, we adopted the kinematic and dynamic parameters provided in one of the value-added catalogues part of DR3. We refer to Section 7.3.3 of \cite{buder2021galah} for details. For the \titan, we followed the same method to compute the orbits. We used the heliocentric velocities and distances given in \cite{giribaldi2021A&A...650A.194G}. Orbits were computed with the {\sf galpy} package \citep{bovy2015ApJS..216...29B} using the gravitational potential of \cite{McMillan2017MNRAS.465...76M}. The orbits were integrated for 13.8 Gyr. For the solar Galactocentric distance and the circular speed at the Sun's position, we adopted the values determined in \cite{McMillan2017MNRAS.465...76M}. For the solar peculiar velocity, we adopt the values given from \cite{schonrich2010MNRAS.403.1829S}. 

\section{Accuracy and consistency of the parameter scales}
\label{sec:consistency}

Atmospheric parameters (\teff, \logg, and [Fe/H]) of the \titan\ are those determined in \cite{giribaldi2021A&A...650A.194G}. For the determination of \teff, the method relies on fitting the observations with synthetic H$\alpha$ line profiles computed using hydrodynamic three-dimensional (3D) model atmospheres and taking into account departures from the local thermodynamic equilibrium \citep{amarsi2018}. As demonstrated in \cite{giribaldi2021A&A...650A.194G}, an accuracy of the order of 1\% was reached for the \teff\, values. To determine \logg, Gaia parallaxes of the early data release 3 \citep{2021A&A...649A...2L}, corrected for the biases discussed in \citet{2021A&A...649A...4L}, were used. A similar accuracy level of $\sim$1\% was achieved. 

\begin{figure}
    \centering
    \includegraphics[width=1\linewidth]{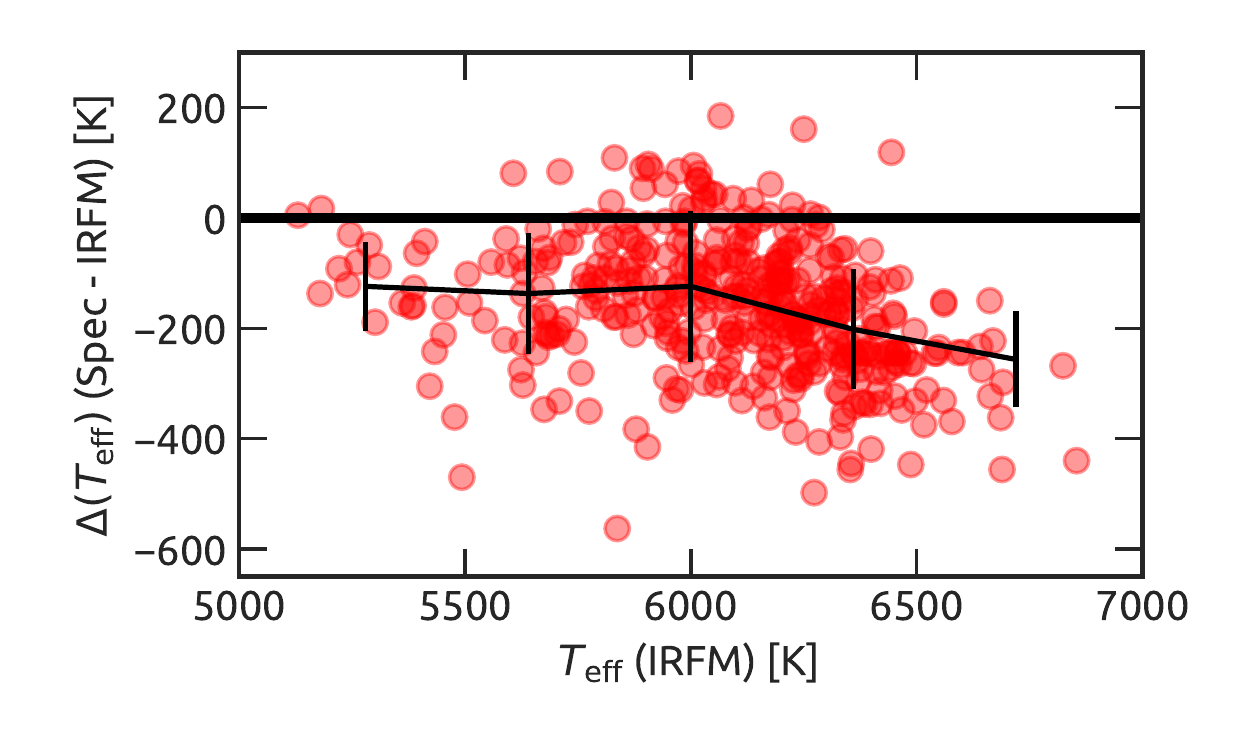}
    \caption{ {\bf Offset between spectroscopic and IRFM \teff\ values.} The distribution includes 408 turnoff stars from the GALAH catalog, restricted according to the selection explained in Sect.~\ref{sec:data}. Only stars with halo-like orbits, i.e. those confined within the retrograde and null net rotation areas in Fig.~\ref{fig:populations}, are shown.
    The thick continuous line connects the medians computed in equally spaced bins of IRFM \teff. Vertical bars indicate the 1$\sigma$ dispersion.
    }
    \label{fig:teff_bias}
\end{figure}

In \cite{giribaldi2021A&A...650A.194G}, it was shown that the accurate \teff\ scale of the \titan\, is compatible with the scale set by the InfraRed Flux Method (IRFM) implementation in \cite{Casagrande2010}.
New IRFM relations using photometry in the Gaia system were recently published by \cite{casagrande2021}, on a scale that was made compatible with that of the previous work. 
Therefore, for our selected sample of GALAH stars, we decided to adopt the IRFM \teff\ of \cite{casagrande2021} instead of the spectroscopic values available in the GALAH DR3 catalogue. 
Fig.~\ref{fig:teff_bias} shows the offset between spectroscopic and IRFM~\teff\ values for our selection of turnoff stars. The plot includes the turnoff stars within the sample of 6250 halo stars pre-selected according to what is explained in Sect.~\ref{sec:data}. Spectroscopic temperatures appear to be underestimated by $\sim$100--200 K.

Furthermore, the values of \logg\ of the GALAH stars were redetermined here,  using the new IRFM \teff\ values, to be consistent with the method used in \cite{giribaldi2021A&A...650A.194G}, i.e. \logg\, was determined by fitting Yonsey-Yale isochrones \citep{kim2002,yi2003} using the code $q^2$ \citep{ramirez2013ApJ...764...78R}. As input, the values of \teff, [Fe/H], $V$ magnitude, and parallax ($\varpi$) are required. In this case, [Fe/H] and $V$ values compiled in the GALAH DR3 catalogue were used. 
Metallicity values were not adjusted because they appear to have negligible offsets compared to the scale of the Gaia benchmark stars \citep{jofre2018RNAAS...2..152J}; see Fig.~6 in~\cite{buder2021galah}. The Gaia benchmark stars metallicity scale is already consistent with the metallicity values of the \titan; see \cite{giribaldi2021A&A...650A.194G}. Our stellar ages and the element abundances required for the separation of the populations are based on the scales of \teff\ and \logg\ determined here.

\section{Stellar ages}
\label{sec:ages}

Stellar ages for the \titan\ and the GALAH sample were determined using Yonsey-Yale isochrones \citep{kim2002,yi2003} and the code $q^2$ \citep{ramirez2013ApJ...764...78R}. As mentioned above, the code requires \teff, [Fe/H], $V$, and $\varpi$, and their corresponding errors as input. 
The code uses the $\varpi$ value to estimate the absolute magnitude $M_V$ using a distance value that comes from a simple parallax inversion. Here, we decided to adopt for our stars the geometric distances from \cite{bailer-jones2021AJ....161..147B}, which have already been corrected for the parallax zero-point offsets. As the code needs parallaxes as input, we simply inverted the distances from \cite{bailer-jones2021AJ....161..147B}. As parallax errors, those directly given in the Gaia catalogue were adopted; for our nearby stars, they are virtually equivalent to those obtained by inverting the 16th and 84th percentiles of the geometric distance posterior. 
For the GALAH sample, $V$ magnitudes were corrected from extinction, using $-3.1\,\times \, E(B-V)$, and adopting reddening values from the GALAH DR3 catalogue. These values were estimated as described in \cite{casagrande2021}. 
The GALAH observations mostly avoid the Galactic plane  \citep[see Fig.~27 in][]{buder2018MNRAS.478.4513B}, which implies small reddening values for most of the sample.
In fact, for 85\% of the stars that we selected, the $E(B-V)$ values are distributed between 0 and 0.1~mag, with a peak at 0.04~mag. In addition, for 10\% of the stars, the $E(B-V)$ values vary between 0.1 and 0.2~mag while the remaining 5\% have $E(B-V) > 0.2$~mag.
For these 5\% (which correspond to eight stars), we compared our adopted reddening values with estimates made using the database \textit{Stilism}\footnote{\url{stilism.obspm.fr}} \citep{Capitanio2017}. We found a systematic offset of about $\sim$0.1~mag that would in turn increase the age of these few stars by about 1~Gyr. We note here that we assumed  $\pm$0.08~mag as the typical combined value of extinction and magnitude uncertainty, thus the effect of such offsets is essentially already included in the error we estimated for the ages. We decided to still use $E(B-V)$ from GALAH for these eight stars, to ensure consistency with the rest of the sample.

\begin{figure}
    \centering
    \includegraphics[width=1.0\linewidth]{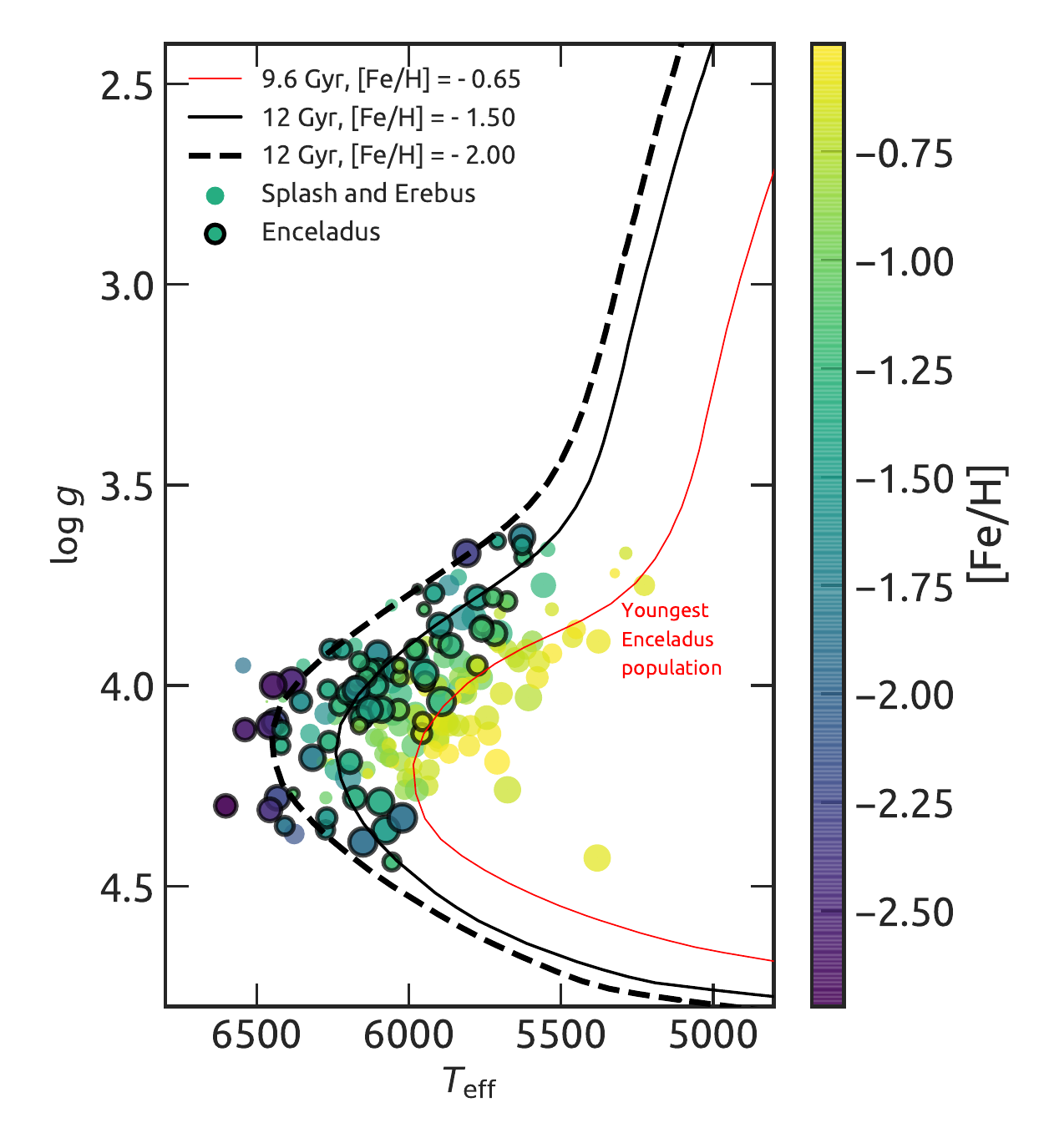}
    \caption{{\bf Kiel diagram with the stars divided into the different stellar populations.} The symbols are color-coded according to metallicity as indicated in the scale on the right side. The symbol size is proportional to the stellar age. Stars belonging to Gaia-Enceladus are highlighted by dark contours and have age precision better than 1.2~Gyr. Stars belonging to the \spl\ and \Ere\ have age precision better than 1.5~Gyr and 2.0~Gyr, respectively. Isochrones are overplotted for reference. The isochrone corresponding to the youngest Gaia-Enceladus population discussed in Sect.~\ref{sec:discusion} is highlighted in red. These stars are the same shown in
    Fig.~\ref{fig:enceladus} and \ref{fig:feh.age}. }
    \label{fig:kiel}
\end{figure}


Turnoff stars were selected for the determination of age because the spacing between the isochrones in this region allows for more precise values to be estimated compared to any other place on the HR diagram (see Fig.~\ref{fig:kiel}). The most important factor that limits the precision in isochronal ages for single stars is the error in \teff; in this work, the average accuracy in \teff\, is 50~K, and this is unmatched in any other stellar sample. Another important factor is the determination of \logg\ or luminosity (closely related quantities). For that, precise parallaxes are needed, such as those that have become available thanks to Gaia EDR3. All these conditions have allowed us to derive ages with an average precision of $6.5\%$ (0.9 Gyr) for 45 of the 48 stars in the \titan\ sample and with an average precision of $10\%$ (1 Gyr) for 183 stars of the GALAH catalogue. For stars in the \spl\ and \Ere\ populations, we accepted ages that are slightly worse to increase their numbers. In such cases, the errors can be up to 1.5 and 2.0 Gyr for \spl\ and \Ere, respectively.

\section{Chemical abundances}
\label{sec:abundances}

To derive chemical abundances for the \titan, we adapted {\sf iSpec} \citep{blanco-cuaresma2014} Python subroutines to be able to perform a supervised line-by-line fitting. Synthetic spectra calculations were performed with the {\sf Turbospectrum} radiative transfer code \citep{turbospectrum} using the MARCS \citep{gustafson2008} model atmospheres. Abundances were derived keeping all stellar parameters fixed to the values obtained in \cite{giribaldi2021A&A...650A.194G}: \teff, \logg, [Fe/H], resolution, projected rotational velocity (\vsini), microturbulence velocity ($v_{mic}$), and macroturbulence velocity ($v_{mac}$). 

We adopted the solar meteoritic abundances of \cite{grevesse2007SSRv..130..105G} as the zero point. The following spectral lines were used to calculate the abundances: Mg I lines at  $\lambda$5528~\AA\ and $\lambda$5711~\AA; Ti I lines at $\lambda4911.19$, $\lambda$4981.73, $\lambda$4999.50, $\lambda$5016.16, and $\lambda$5381.02~\AA; Si I lines at $\lambda$5684.48, $\lambda$5708.40, and 6371.37~\AA; Ni I lines $\lambda5035.362$ and $\lambda5476.904$~\AA; Eu II lines at $\lambda4129.72$ and $\lambda4205.03$~\AA; and Ba II lines at $\lambda5853.67$, $\lambda6141.71$, $\lambda6496.90$~\AA. We used the atomic data from the Gaia-ESO line list \citep{heiter2021A&A...645A.106H}.

The abundances of Mg include corrections for atomic diffusion, which we obtained by interpolating the values of \cite{korn2007} and \cite{gruyters2013}. These corrections only affect stars with metallicities below [Fe/H] = $-1.65$~dex. At higher values of [Fe/H], the corrections for Mg cancel out with those for Fe. For stars with [Fe/H] below $-$2.10, the correction is of $+0.1$ in [Mg/Fe] and [Si/Fe]. The correction decreases linearly between [Fe/H] = $-2.10$ and $-$1.65, where they become zero.

\begin{figure}
    \centering
    \includegraphics[width=1.0\linewidth]{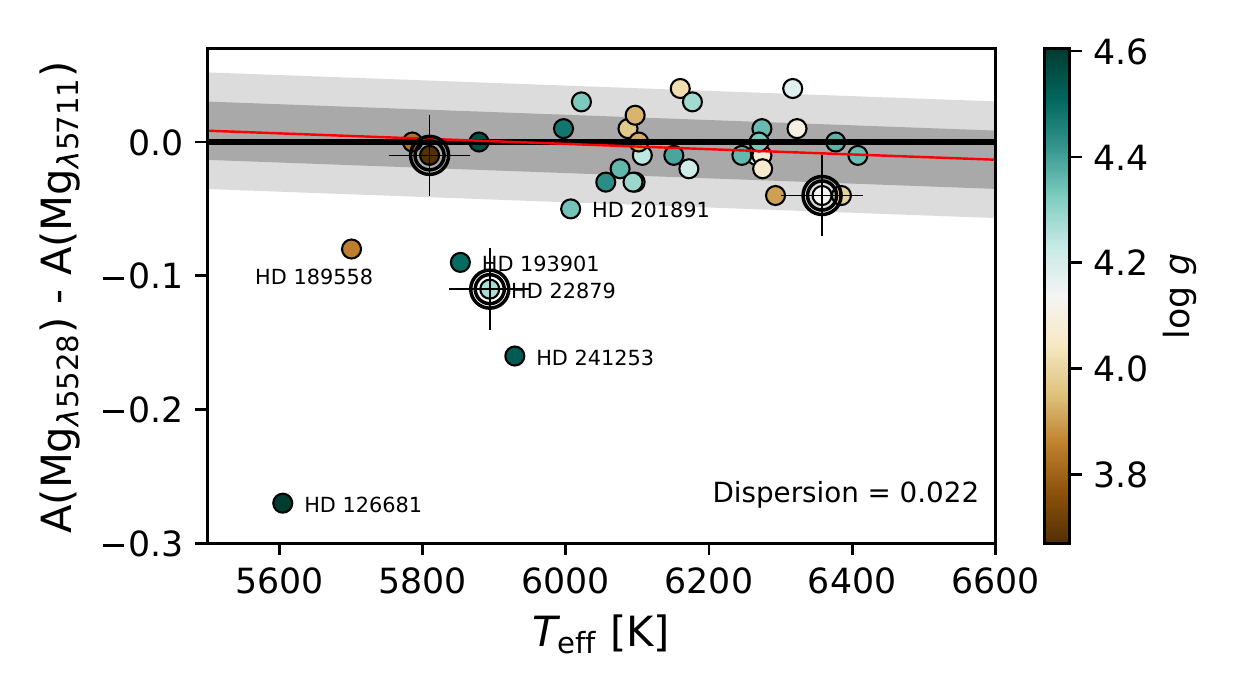}
    \includegraphics[width=1.0\linewidth]{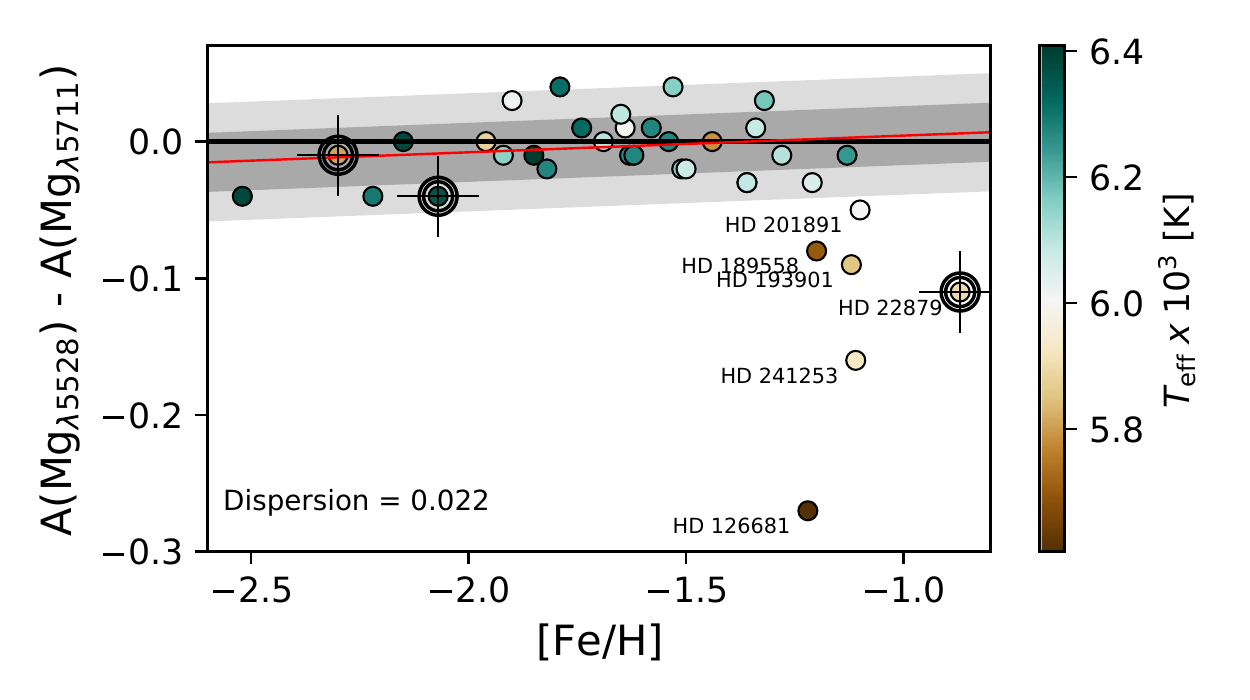}
    \includegraphics[width=1.0\linewidth]{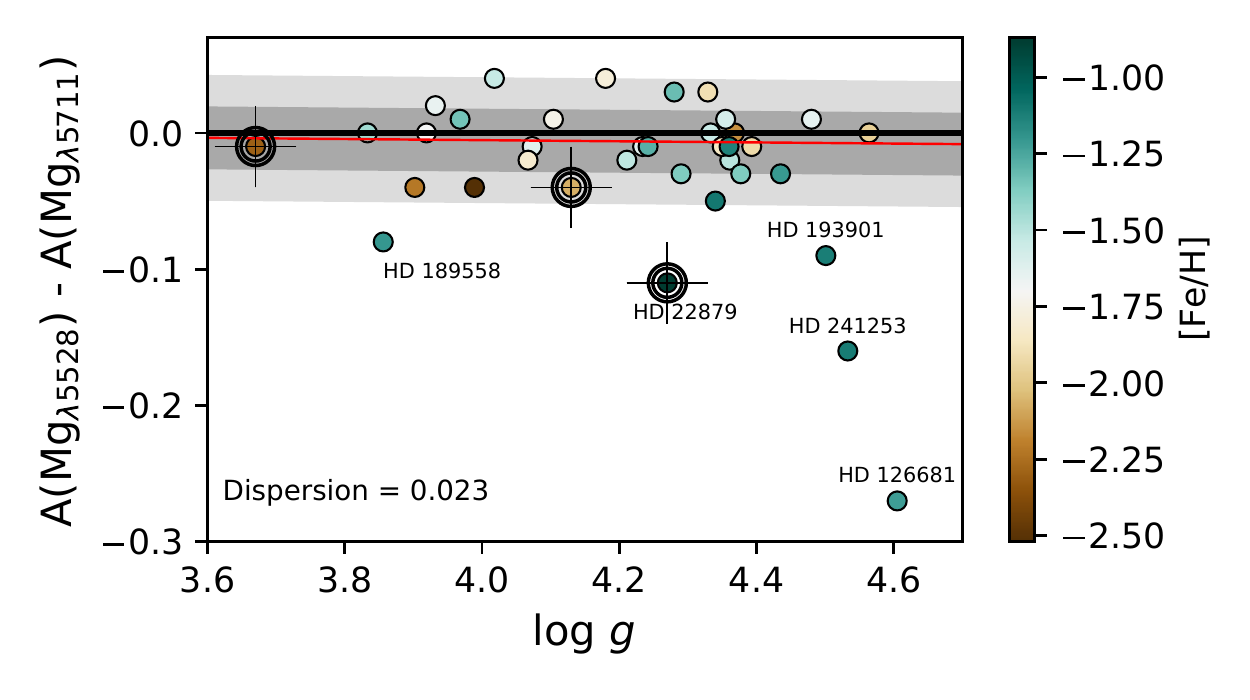}
    \caption{\textbf{Consistency of Mg abundance from the lines 5528 and 5711~\AA.} For the \titan, difference between Mg abundances from the lines $\lambda$5528~\AA\ and $\lambda$5711~\AA\ as functions of the atmospheric parameters. 
    The red lines are the linear regressions and the shaded areas indicate 1$\sigma$ and 2$\sigma$ dispersions. 
    The reference stars HD~22879, HD~140283, and HD~84937 are identified with bold contours.}
    \label{fig:Mg_comparison}
\end{figure}

\subsection{Non-LTE corrections}

Our abundances are computed using one-dimensional model atmospheres and assuming local thermodynamic equilibrium (LTE). However, and in particular for metal-poor stars, deviations from LTE (the so-called non-LTE effects) have been shown to be of importance. Therefore, when possible, we used literature calculations of non-LTE effects to correct the abundance that we derived.

Regarding the abundances of Mg, it was found in \cite{mashonkina2013} that, for turnoff stars (\teff~= 6000~K and \logg~= 4~dex) with metallicities of [Fe/H]~= $-1, -2,$ and $-3$~dex, a constant correction in the abundances of $\sim$+0.05 is needed for the Mg lines analysed here. This is consistent with the fact that we find agreement between the abundances derived from these lines, as we can see in Fig.~\ref{fig:Mg_comparison}. The most extreme outliers in this figure are among the coolest and most metal-rich \titan\ ([Fe/H] $> -1.2$~dex) with the highest \logg. Stars in this range of surface gravity were not considered for the main discussion in the \titan\ or GALAH samples. Therefore, we adopted the constant non-LTE correction of +0.05 from \cite{mashonkina2013} for the Mg abundances of the entire sample. 

For titanium, the abundances obtained from line $\lambda4911$~\AA\ have been found to be affected by negligible non-LTE effects \citep{bergemann2011}. Therefore, we used this line as the main indicator of the Ti abundances. For the other lines, small offsets as a function of [Fe/H] were applied to correct all other lines on the scale defined by the $\lambda4911$~\AA\ lines; see Fig.~\ref{fig:Ti_comparison}. The final abundance of each star was then computed by averaging the values.

For silicon, all the lines used here seem to be free from non-LTE effects \citep{shi2009,shi2011}. Therefore, the final abundances are given by averaging the line values. Figure~\ref{fig:Si_comparison} shows a comparison among the abundances of each line. For Ni, on the other hand, we did not find studies on non-LTE and 3D effects for the lines we used in this work. No correction could be applied in this case. The final Ni abundances are the averages of the line values. Figure~\ref{fig:Ni_comparison} compares the abundances of both lines used here.

For calcium, we extracted the non-LTE corrections \citep{mashonkina2007} for the lines lines $\lambda$6166 and $\lambda$6169 \AA\ by running the NLTE MPIA online tool \citep{NLTE_MPIA}. For the line line $\lambda$6166~\AA, the non-LTE corrections are zero for all combination of atmospheric parameters. For the line $\lambda$6169~\AA, the corrections vary with \teff\ and [Fe/H]. Therefore, we chose the abundances of $\lambda$6166~\AA\ as the standard reference to examine the behaviour of the relative abundances of the other three lines. Figure~\ref{fig:Ca_comparison} shows a comparison of the abundances obtained with these lines. Similarly to what was done above for Mg and Ti, the final Ca abundances were obtained by averaging the offset-corrected values from each line.

For Eu, the work done in \cite{zhao2016ApJ...833..225Z} showed that the non-LTE effects on abundances computed using the line at $\lambda4129$~\AA\ are negligible in metal-poor turnoff stars. Therefore, we used the abundances from this line as a reference and corrected for the relative offsets of abundances from $\lambda4205$~\AA\ as function of [Fe/H]. Figure~\ref{fig:Eu_comparison} compares the abundances of both lines.

For Ba, the work of \cite{mashonkina1999A&A...343..519M} shows that the abundances determined from the line at $\lambda5853$~\AA\ are not affected by non-LTE effects in metal-poor turnoff stars. Therefore, we used the abundances of this line as a reference to correct the relative offsets of the other lines as a function of [Fe/H]. Figure~\ref{fig:Ba_comparison} compares the abundances of both lines.
The final abundances are then given by the averages. 

Figure~\ref{fig:other_elements} shows plots with the abundances of the elements above relative to hydrogen as a function of stellar age, including only the \titan. Only the Enceladus and \Ere\ populations (defined in Sect.~\ref{sec:separation}) are distinguished because stars from other populations were not found in this sample. The main discussion in Sect.~\ref{sec:discusion} uses mostly the abundances of magnesium.

\subsection{Uncertainties of the abundances}

The uncertainties of the abundances were computed by estimating the separated effect of each atmospheric parameter. The typical uncertainties of the \titan\ atmospheric parameters are $\pm$40~K in \teff, $\pm$0.04~dex in \logg, $\pm$0.05~dex in [Fe/H], and $\pm$0.1~km s$^{-1}$ in $v_{mic}$. New abundances are computed for each spectral line by changing each parameter, one at a time, by its typical error. The influence of errors in $v_{mic}$ was found to be negligible. To estimate the final uncertainty, all changes in the abundances are then added quadratically. Finally, to that uncertainty estimate, we also add the standard deviation of the abundances of the multiple lines. For Mg, for example, this is of $\pm$ 0.023~dex, as shown in Fig.~\ref{fig:Mg_comparison}. The typical uncertainties are 0.037, 0.023, 0.029, 0.031, 0.041, 0.080, and 0.041 for A(Mg), A(Si), A(Ca), A(Ti), A(Ni), A(Ba), and A(Eu), respectively.

\subsection{Magnesium abundances of the GALAH stars}

Corrections were applied to the Mg abundances of the GALAH stars to ensure consistency with the abundance scale of the \titan. For that, we used the offset between the new values of \teff\ and \logg\ that we adopted here and the original values from GALAH DR3. As we mentioned before, the values of [Fe/H] did not change, so there is no offset in this parameter. The offsets in the parameters were mapped into abundance offsets using the same approach adopted to calculate the abundance uncertainties. We found that the effect of the \logg\ offset in Mg abundance is negligible. Using the line $\lambda5711$~\AA\ for this analysis, we found that the necessary change in the Mg abundances from GALAH is an offset of +0.07 dex, independent of atmospheric parameters. The fact that the correction is just a constant offset guarantees that the change in [Mg/Fe] values along the age-metallicity relation discussed in Sect.~\ref{sec:discusion} (Fig.~\ref{fig:enceladus}) is not an artefact of our abundance scale correction. The GALAH Mg abundances already take into account non-LTE effects, so this type of correction does not need to be applied again.

\section{Separation of stellar populations} 
\label{sec:separation}

\begin{figure}
    \includegraphics[width=0.95\linewidth]{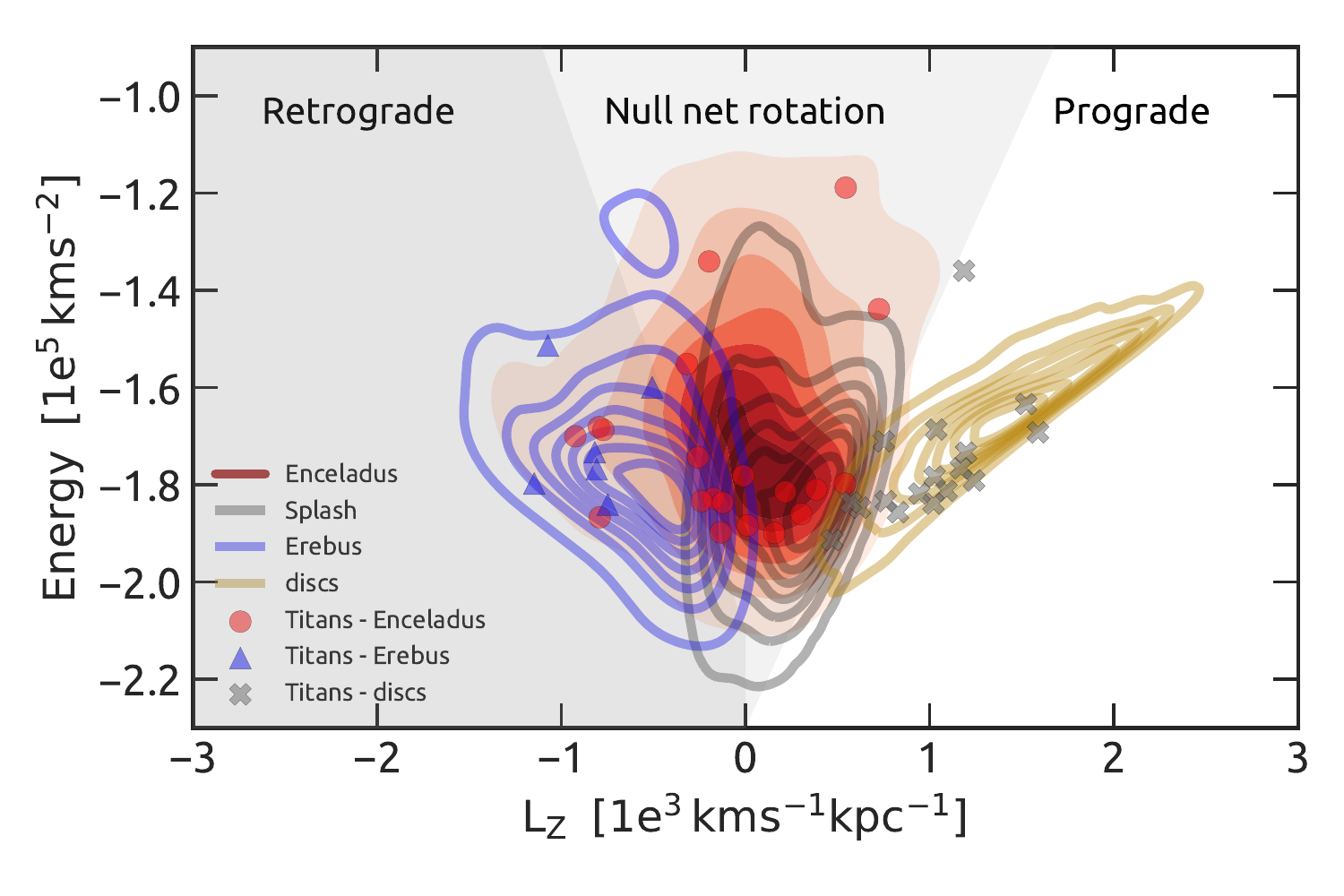}
    \includegraphics[width=0.95\linewidth]{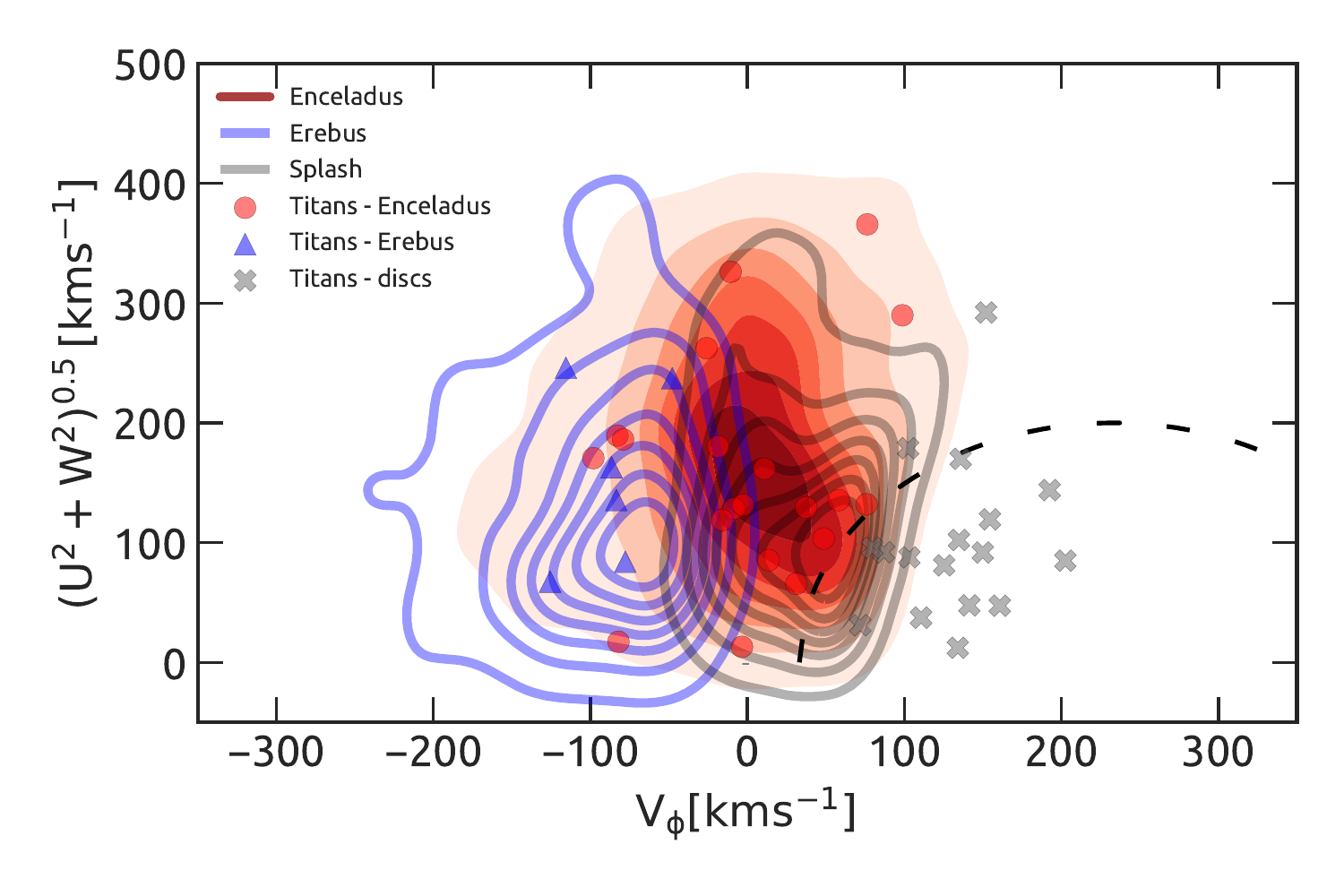}
    \includegraphics[width=0.95\linewidth]{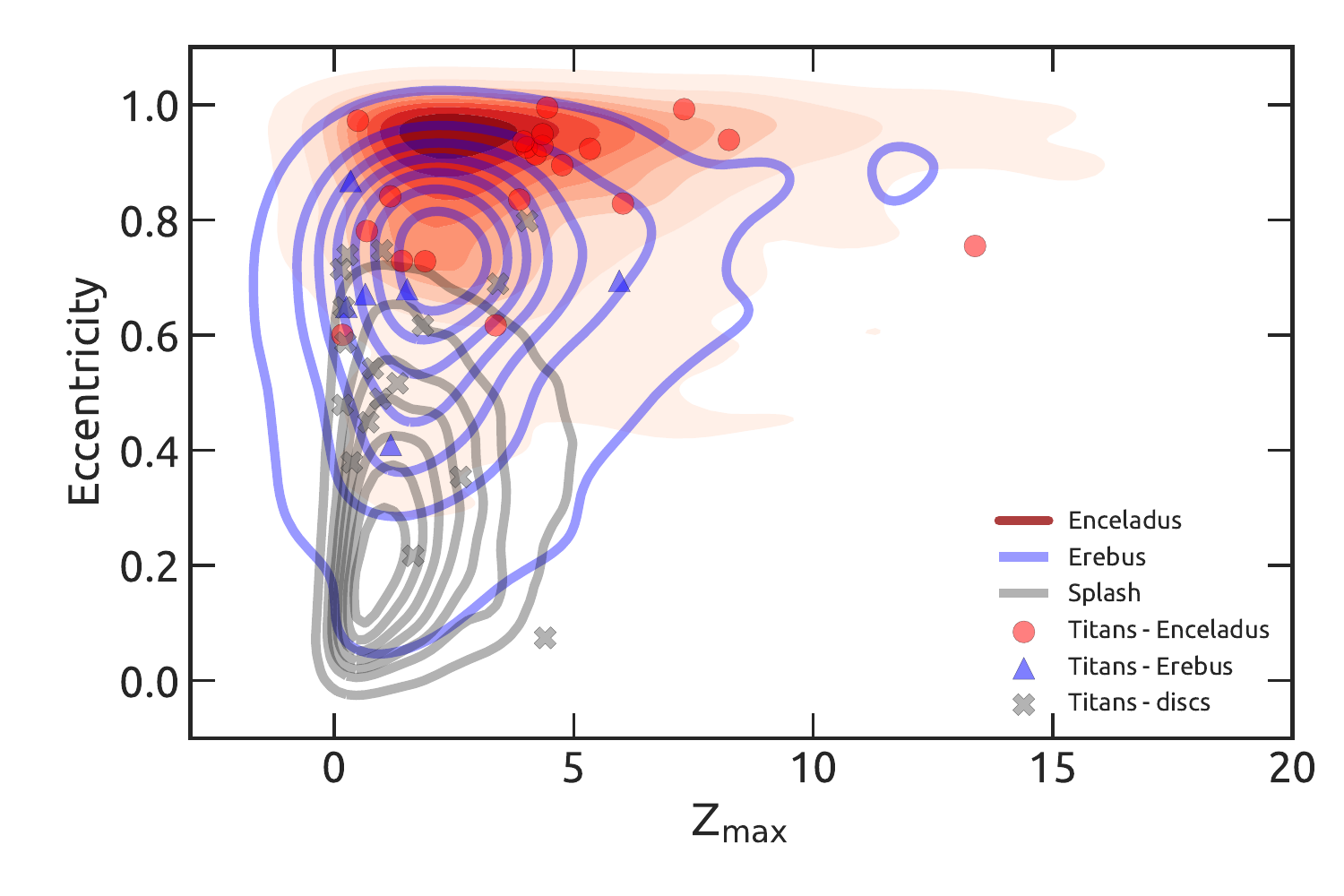}
    \caption{\textbf{Inner halo populations in kinematic and dynamical diagrams.} \textit{Top panel:} Lindblad diagram. The contours represent the 33, 66, 90, 95, 98, 99, and 99.9~\% cumulative distribution of the populations disentangled with the \titan\ and GALAH metal-poor stars. The regions and stars corresponding to the Gaia-Enceladus, \spl, and \Ere\ populations are highlighted. Here, the \titan\ that have disc dynamics are also shown for completeness. \textit{Mid panel:} Toomre diagram. Symbols and colors are the same as in the top panel. The region of disc stars, selected following \cite{helmi2018}, is delimited by the dashed line. \textit{Bottom panel:} Eccentricity vs. maximum distance from the Galactic plane ($Z_{max}$). Symbols and colors are the same as in the other panels.}
    \label{fig:populations}
\end{figure}

Our goal here is to classify the combined (GALAH + \titan) sample of stars into those stars that have an accreted origin and those that most likely formed in situ. As a first step to separate the metal-poor stellar populations, we used the Toomre diagram to select and exclude stars with prograde motion and total velocity below 210 km s$^{-1}$ (see Fig. \ref{fig:populations}). Most stars with total velocities lower than that are probably part of the Milky Way disc. In addition, the 
stars Wolf 1492 (G64-12), BPS CS 22166-30, and CD-71~1234 (the first two are the most metal-poor stars in the sample) are the only \titan\ to reach maximum distances from the Galactic plane above 20 kpc.
They have been considered to be part of the outer halo population and, as such, are not used in our main discussion. The remaining \titan make up the sample from which we have the highest probability of identifying stars originating from, in particular, the Gaia-Enceladus merger.

Although there is no certain way to identify stars of accreted origin, there are a few telltale signs that one can use. In particular, previous mergers are thought to be the main source of stars that are usually found in loosely bound orbits that can be retrograde, found to have high spatial velocities, and/or found to be metal-poor, but display a low [$\alpha$/Fe] ratio \citep{Majewski1992,Gratton2003}. To first order, our classification is based on chemical differences; that is, we consider stars with a lower [Mg/Fe] ratio, at a given metallicity, to have a higher chance of have been accreted during a merger. This separation is then supported and improved by verifying the orbital and kinematic signatures of the accreted populations. In this procedure, the \titan\ and GALAH samples were combined and used as a single sample.

The classification of the populations was implemented using a Gaussian mixture model clustering algorithm \citep[scikit-learn,][]{scikit-learn} and proceeded as follows. First, the space made of angular momentum (L$_{Z}$) and orbital energy ($E$), the so-called Lindblad diagram, was divided into many overlapping sections of dimensions of 0.5 [10$^3$ km s$^{-1}$ kpc$^{-1}$] $\times$ 0.2 [10$^{5}$ km s$^{-2}$] in $\Delta L_{Z}\, \times \, \Delta$Energy. Each region was examined separately to find the dominant pattern in the [Mg/Fe] vs.\ [Fe/H] space. Mixtures of two, three, and four clusters were tested by visual inspection. In general, a configuration with three clusters seemed to better capture the division between two groups with low and high [Mg/Fe] and a third group of stars with uncertain classification.

\begin{figure}
    \centering
    \includegraphics[width=1\linewidth]{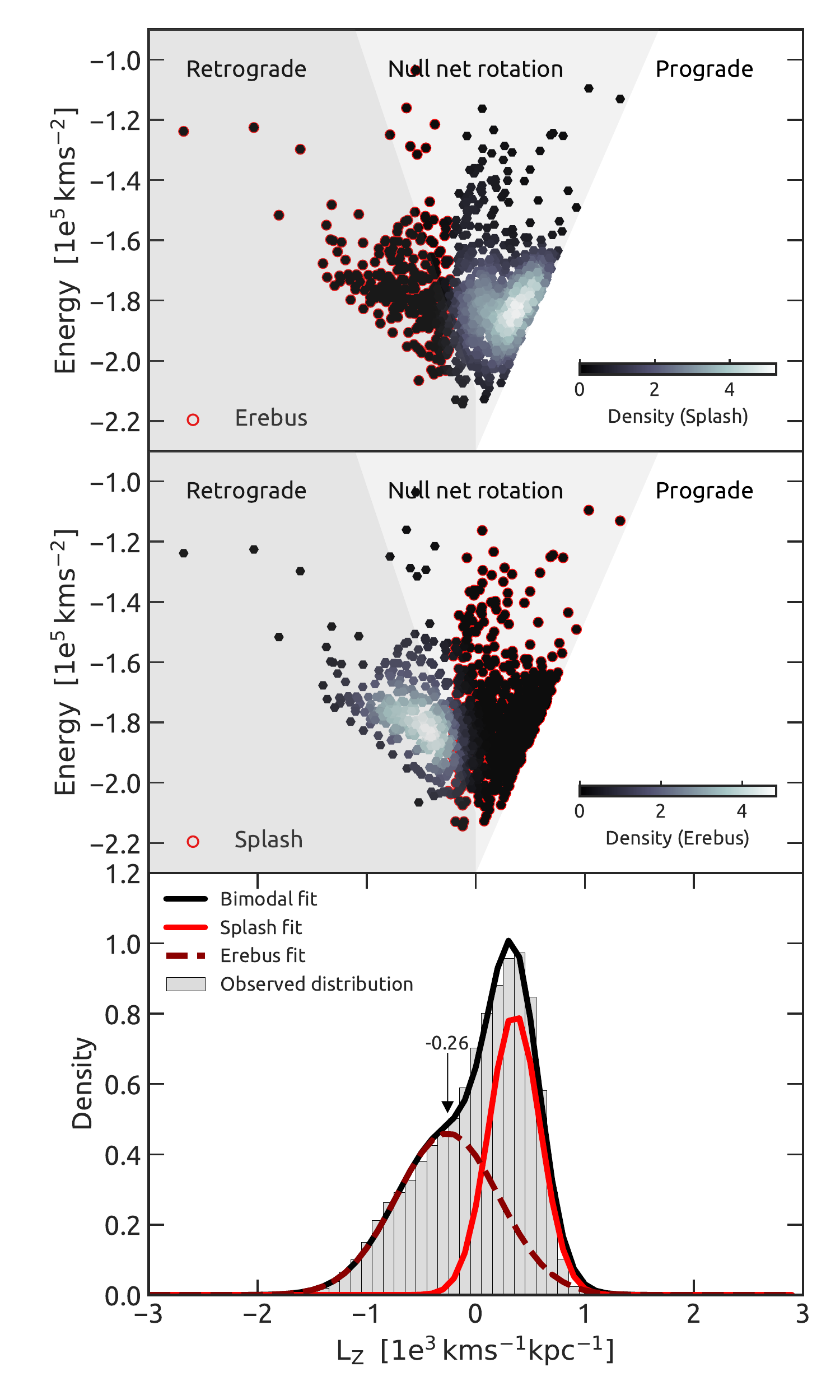}
    \caption{{\bf Lindblad diagram of the \spl\ and \Ere\ populations.} Top and mid panels highlight the probability density of the \spl\ and \Ere\ stars, respectively. The bottom panel shows the histogram of angular momentum, $L_Z$, for stars with binding energy between $-1.7$ and $-1.9 \times 10^5$ km$^{-2}$, approximately where most stars are concentrated. The histogram is well fit by a bi-modal Gaussian distribution (in black). One peak is dominated by \spl\ stars (solid bright red line) and the other by \Ere\ (dashed dark red line).}
    \label{fig:Splash_thamnos}
\end{figure}
\begin{figure}
    \centering
    \includegraphics[width=1.0\linewidth]{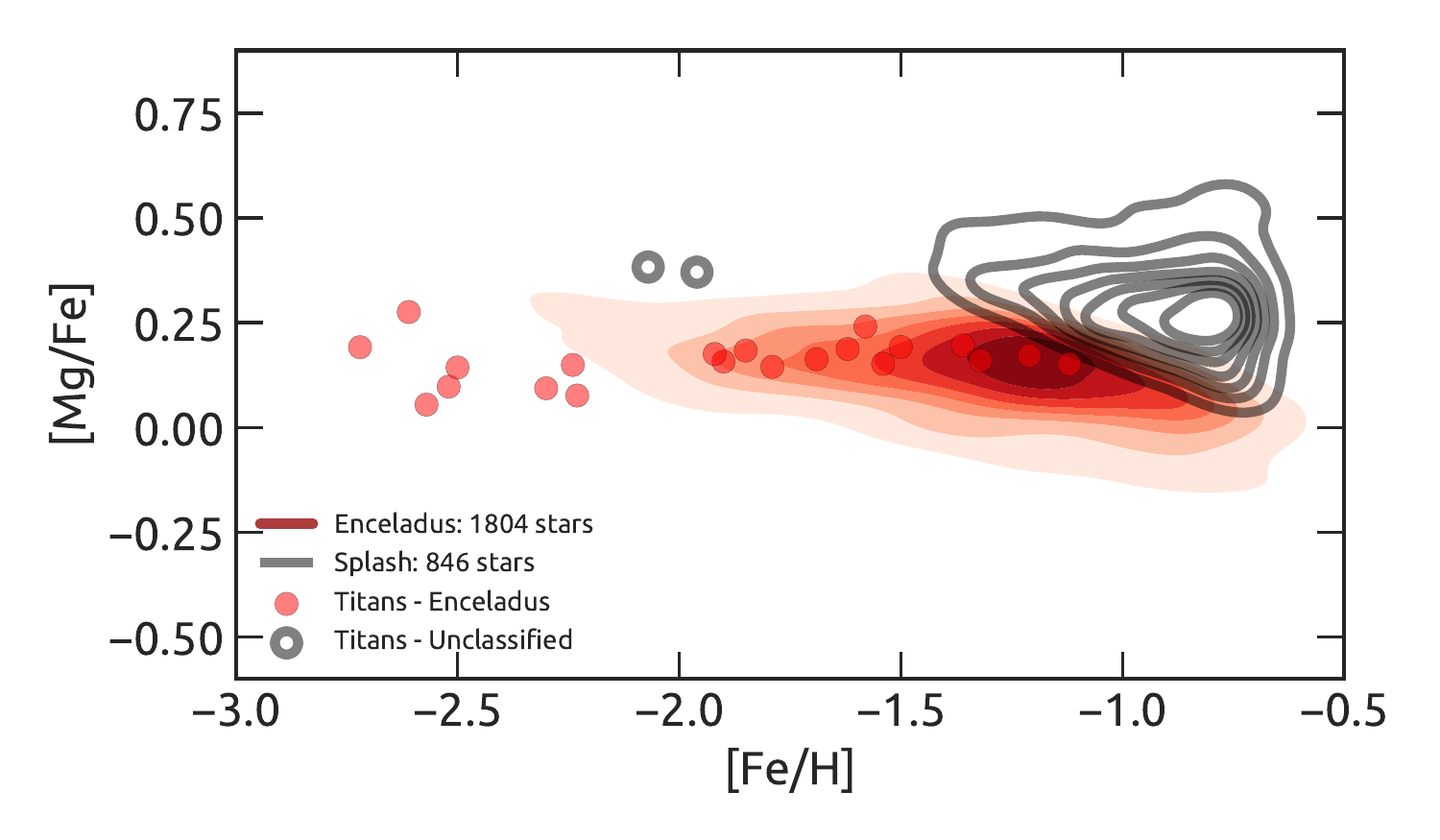}
    \includegraphics[width=1.0\linewidth]{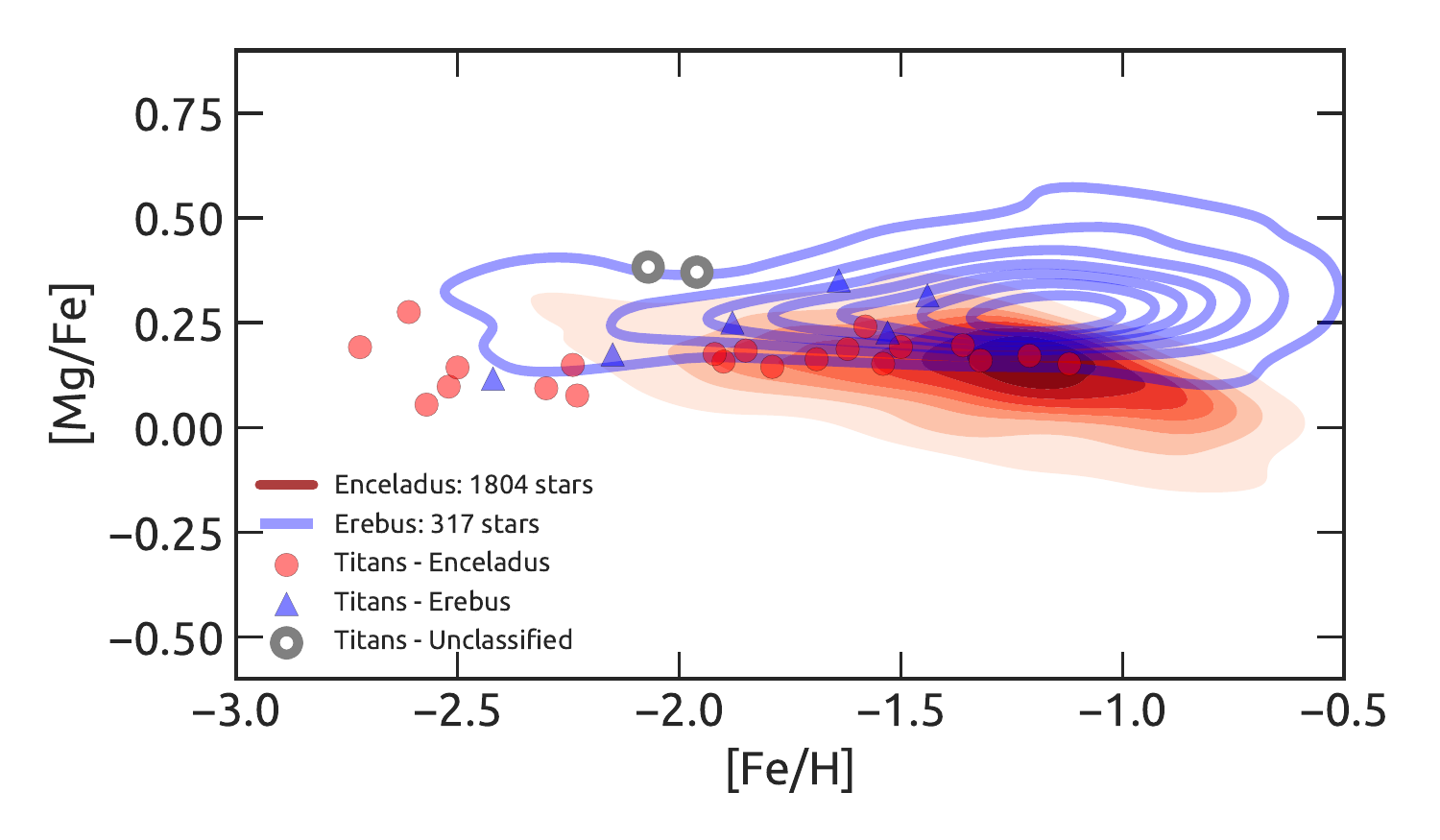}
    \caption{\textbf{Inner halo populations in the [Mg/Fe] vs. [Fe/H] diagram.} Cumulative distribution plots of the populations selected with GALAH in the [Mg/Fe] vs. [Fe/H] space. The contour colors are consistent with those used in Fig.~\ref{fig:populations}. The symbols display the \titan\ associated to Gaia-Enceladus and \Ere, as well as two unclassified stars. \textit{Top panel:} The regions occupied by the stars from Gaia-Enceladus and the \spl. \textit{Bottom panel:} The regions occupied by the stars from Gaia-Enceladus and \Ere.
    }
    \label{fig:sanity}
\end{figure}

From the literature \citep[e.g.,][]{helmi2018,feuillet2021MNRAS.508.1489F}, it is known that retrograde stars tend to have higher [Mg/Fe] ratios and thus are separated from Gaia-Enceladus stars in the chemical diagram. Cone-shaped areas in the Lindblad diagram were tested to determine the $L_Z$ limits that correspond to such a chemical division. In this exercise, it was found that retrograde stars with high [Mg/Fe] extend to very low [Fe/H] values; this is the population we call \Ere, a name chosen for the reasons explained in Sect.~\ref{sec:discusion}. On the other hand, stars that have null net rotation and high [Mg/Fe] mostly have high values of [Fe/H]; this is the population that we identify with the \spl. The division between \Ere\ and \spl\ is further refined as follows. First, we removed the low [Mg/Fe] stars, identified with Gaia-Enceladus, from the sample. Then, we examined the area that the remaining stars occupy in the Lindblad diagram (Fig.~\ref{fig:Splash_thamnos}), where a clustering of two groups was applied. The top and mid panels of the figure show the classification obtained and, in grey scale, the probability density distribution of the groups. The bottom panel of the figure shows that the distribution in terms of angular momentum is clearly bimodal and well-fit by two Gaussians. The peak at prograde orbits is dominated by \spl\ stars, while the peak at slightly retrograde orbits, $L_Z = -0.26\times10^3$ km s$^{-1}$ kpc$^{-1}$, is dominated by the stars that we associate with \Ere. This value is likely displaced from zero to slightly negative because of the initial cut of stars in the prograde area where mostly disc stars remain. This shows that the two populations extend beyond the limits set by the clustering algorithm. On the other hand, it is evident that each group is certainly dominated by stars with their own characteristic Galactic rotation pattern. The final selection of the Gaia-Enceladus, \Ere, and \spl\ stars is shown in red, blue, and grey contours, respectively, in the diagrams of Fig.~\ref{fig:populations} and \ref{fig:sanity}. While the distribution of dynamic properties of these groups overlaps somewhat, in the next Section we show how the different stellar groups have well-separated relations in age-[Fe/H]-[Mg/Fe] diagrams. Separation in these age-chemical relations shows that our selection criteria successfully group stars of common origin, perhaps with only a small fraction of cross-contamination among them.

Four of the \titan, BD+26~2621 (11.9~Gyr), HD~74000 (9.5~Gyr), Ross~453 (14.6~Gyr), and HD~106038 (13.8~Gyr), are classified as stars from Gaia-Enceladus, but are located in the retrograde area of the Lindblad diagram. These four stars also have the lowest eccentricity values among the \titan\ (0.6 to 0.7, see the lower panel in Fig.~\ref{fig:populations}). Although Gaia-Enceladus stars are usually assumed to have high eccentricities \citep{Naidu2020}, it has also been suggested that its stars can have a range of eccentricity values \citep{Perottoni22}. In our analysis, what classifies these four stars as members of Gaia-Enceladus is their low [Mg/Fe] ratio. Except for HD~74000, they are old stars born within the first two billion years after the Big Bang. HD~74000, although with relatively young age in this sample, has a low metallicity ([Fe/H] $= -1.85$), which locates it to the right of the main Gaia-Enceladus sequence in the age-metallicity plot of Fig.~\ref{fig:feh.age}. Due to its age, it seems to have a low probability of belonging to the population that we identified as \Ere, which is mainly composed of stars located to the left of the Gaia-Enceladus sequence (see Fig.~\ref{fig:feh.age}). BD+26~2621 and Ross~453 have low metallicities ([Fe/H] $= -2.42$ and [Fe/H] $= -1.92$, respectively), which locate them in a region of the age-metallicity diagram where the Gaia-Enceladus and \Ere\ sequences are mixed. The correct population to which they belong can be discussed, but removing them from the Gaia-Enceladus sample would not change our results. HD~106038, on the other hand, indeed seems to be an ambiguous case. Its [Mg/Fe] and [Fe/H] values put this star inside the region of highest probability of Gaia-Enceladus stars in the [Mg/Fe] vs.\ [Fe/H] plot (Fig.\ref{fig:sanity}). However, its age seems to be in better agreement with the path of the \Ere\ [Fe/H]-age sequence in Fig.\ref{fig:feh.age}. HD~106038 is known to be a chemically peculiar star that likely underwent the influence of a nearby hypernova \citep{2008MNRAS.385L..93S}. Most likely, its position in the age-metallicity diagram cannot be trusted. In any case, we remark that removing the four stars from the Gaia-Enceladus population does not change our conclusions. The remaining stars in Fig.~\ref{fig:feh.age} would still define the same age-metallicity sequence for Gaia-Enceladus. This discussion just highlights that, although we cannot be sure how to assign the population to each individual star, the general properties of the populations that we defined seem to be robust.

\section{Discussion and conclusions}
\label{sec:discusion}

Distinguishing between metal-poor stars that have been formed in accreted systems or in different in-situ populations is by no means straightforward \citep{Jean-Baptiste2017,Pagnini2022}. Stars considered to have a high probability of having an accreted origin are usually those found in loosely bound orbits that can be retrograde, found to have high spatial velocities, and/or those that are relatively metal-poor, but display a low [$\alpha$/Fe] ratio \citep{Majewski1992,Gratton2003}. However, there is no single criterion that can determine the stellar origin with full certainty. To assign our sample stars to a given stellar population or to a given accreted system, we make use of combined information coming from kinematics, orbits, and elemental abundances. We believe to have reached a final classification of high confidence, where even if some individual cases are incorrect, our conclusions are not strongly affected. Taking into account the 45 \titan, with precise ages, we assign 17 stars to the Milky Way disc, 20 stars to Gaia-Enceladus, three stars to the outer halo, and five stars to an inner halo of probable in situ origin (\Ere, see below). The GALAH sample with precise ages has 183 stars. We assign 61 of them to Gaia-Enceladus, 92 to the \spl, and 30 to \Ere. These stars classified into different populations are shown in kinematic, dynamical, and chemical diagrams in Figs.~\ref{fig:populations}~and~\ref{fig:sanity}.

\begin{figure}
    \centering
    \includegraphics[width=0.9\linewidth]{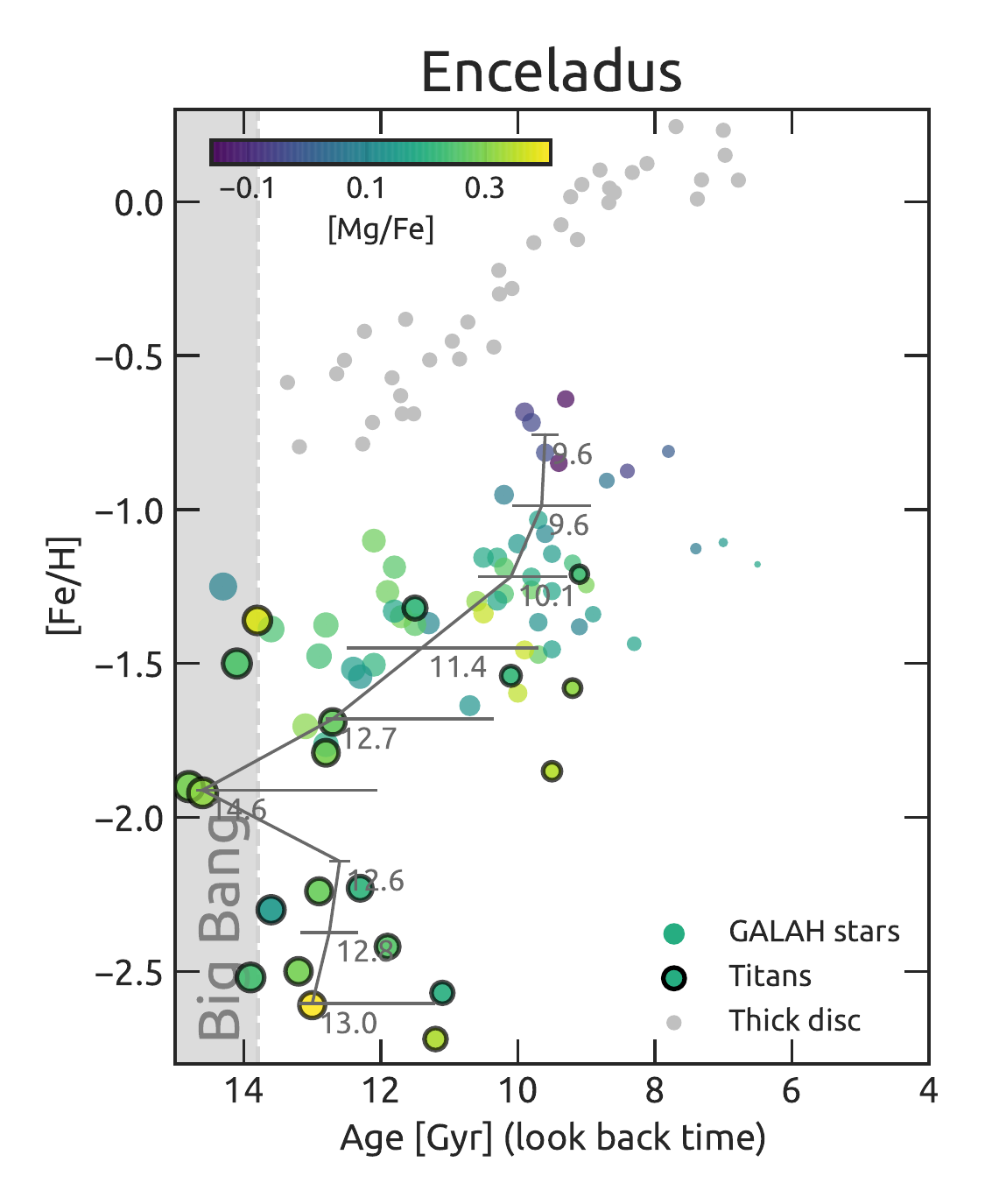}
    \includegraphics[width=0.9\linewidth]{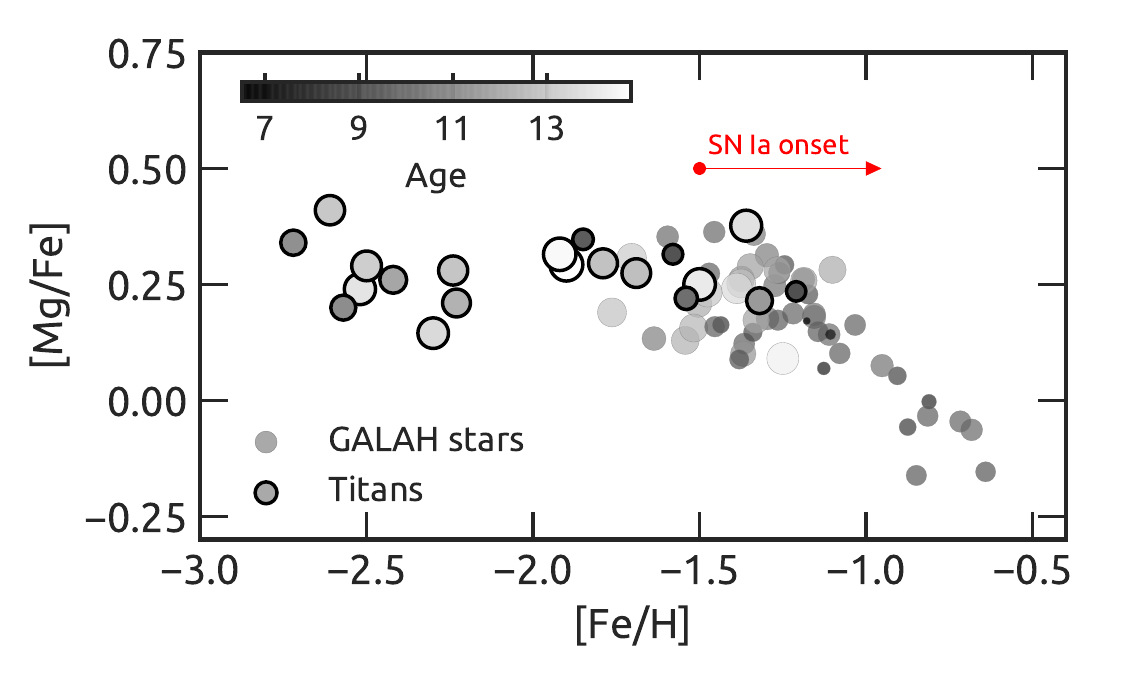}
    \caption{{\bf Chemical evolution of Gaia-Enceladus.} 
    \textit{Top panel:} Time evolution of the chemical enrichment of stars from Gaia-Enceladus. The symbol size changes as a function of stellar age and the points are color-coded according to [Mg/Fe]. Only stars for which an age precision better than $\pm1.2$~Gyr was obtained are shown. The gray line connects median ages computed in equally separated bins of [Fe/H]. The bars correspond to the 25 and 75\% quartiles around the medians. The \titan\ are distinguished from the GALAH stars by the dark contours. Thick disc stars from \cite{haywood2018ApJ...863..113H} are shown as gray circles. \textit{Bottom panel:} The [Mg/Fe] vs. [Fe/H] diagram. The circles are color- and size-coded by stellar age. The approximate metallicity value where Fe coming from supernovae type Ia becomes important is indicated at [Fe/H] $= -1.5$.}
    \label{fig:enceladus}
\end{figure}

\begin{figure*}
    \centering
    \includegraphics[width=0.32\linewidth]{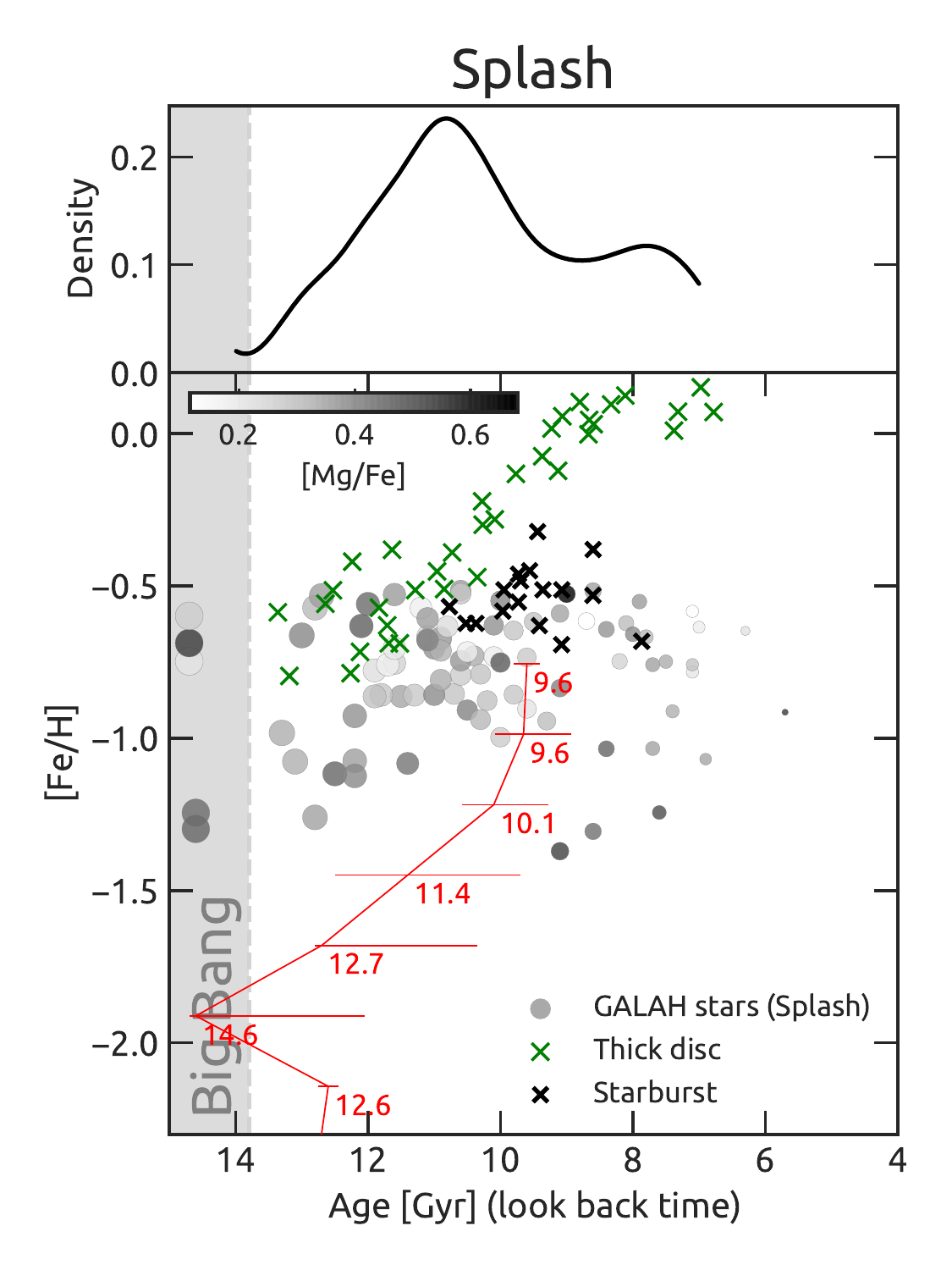}
    \includegraphics[width=0.32\linewidth]{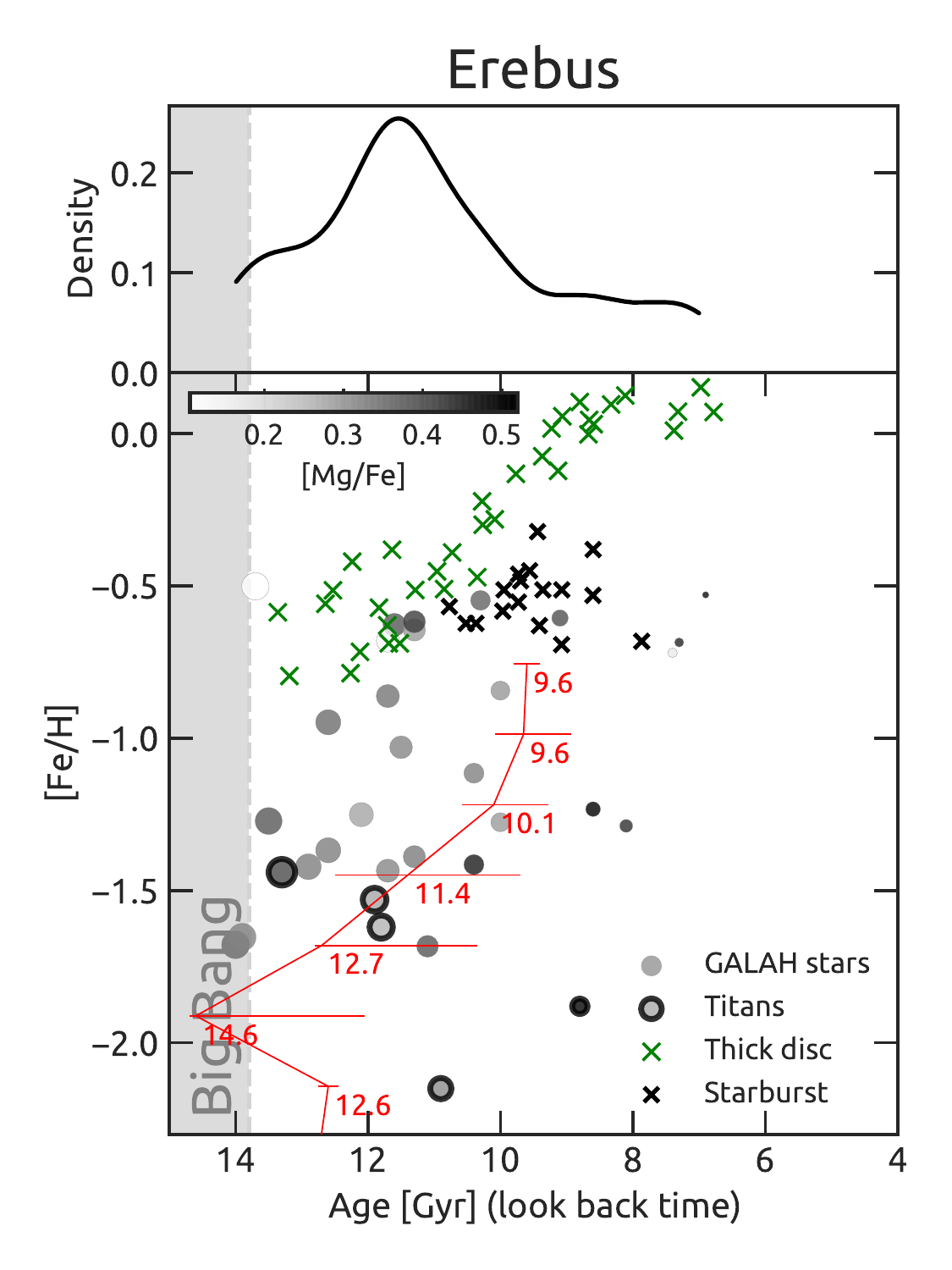}\\
    \includegraphics[width=0.32\linewidth]{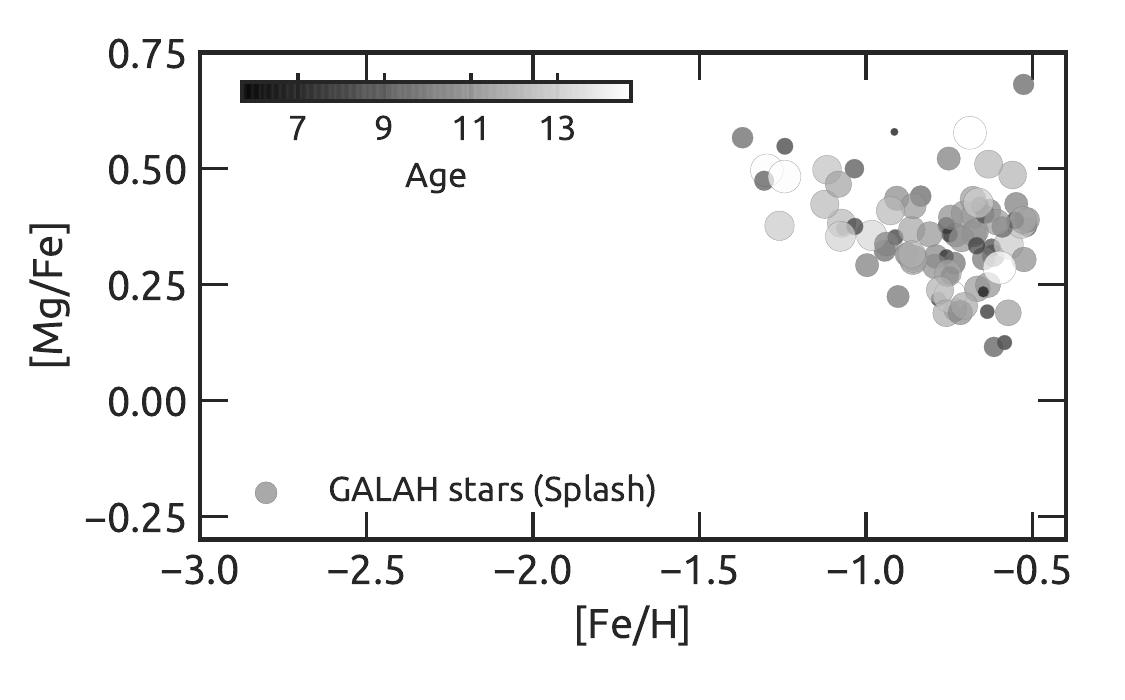}
    \includegraphics[width=0.32\linewidth]{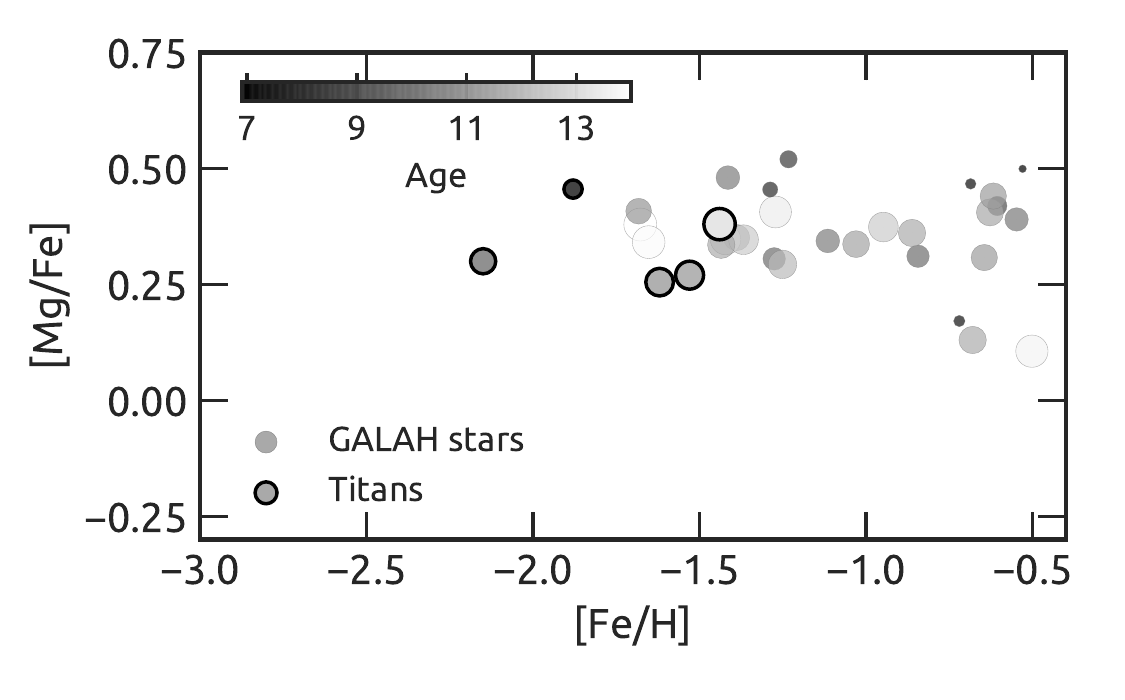}
    \caption{\textbf{Chemical evolution of the \spl\ and \Ere.} 
    \textit{Top:} Age-metallicity relation for stars in the \spl\ (left) and \Ere\ (right) components. The size of the points change according to stellar age and the color according to the [Mg/Fe] ratio. The Gaia-Enceladus sequence is shown as the red line. For the \spl, only stars with ages that have precision better than $\pm1.5$~Gyr are considered. For \Ere, we plot those with age precision below $\pm2$~Gyr. Thick and thin disc stars from \cite{haywood2018ApJ...863..113H} are shown as green and black crosses, respectively. \textit{Bottom:} The [Mg/Fe] ratio as a function of metallicity for stars in the \spl\ (left) and \Ere\ (right) components. Points are color- and size-coded according to the stellar age.}
    \label{fig:feh.age}
\end{figure*}

Figure~\ref{fig:enceladus} shows the chemical enrichment sequence of the Gaia-Enceladus galaxy. This clear chronological step-by-step enrichment of Gaia-Enceladus is revealed here for the first time, thanks to the high precision of the stellar ages. It can be seen that some dispersion remains on the age axis. This can be partially ascribed to uncertainties, which are lower than $\pm0.9$ for the \titan\ and lower than $\pm1.2$~Gyr for GALAH stars. In addition, part of the dispersion is likely intrinsic. It is caused by the stochastic character of chemical evolution in spatially confined groups, which keeps the interstellar medium inhomogeneous at early times \citep{parizot2004A&A...424..747P}.

The [Mg/Fe] ratio remains elevated in stars of Gaia-Enceladus up to an age of $11.3$~Gyr ([Fe/H] $\sim$ $-$1.5 dex). At this point, the ejecta from supernovae (SN) type Ia became important, and the [Mg/Fe] ratio started to decrease. The metallicity where this so-called ``knee'' occurs in Gaia-Enceladus is smaller by 6-7 times in comparison to the metallicity of Galactic thick disc stars where the same feature is seen ([Fe/H]~$\sim-0.7$, grey circles in the top panel of Fig.~\ref{fig:enceladus}). The age we find for the ``knee'' differs from that of the Gaia-Enceladus chemical evolution model presented by \cite{vincenzo2019MNRAS.487L..47V} ($\sim$ 13.5 Gyr). Our result indicates that a revision of this model is in order. 

Figure~\ref{fig:enceladus} shows that the chemical evolution of Gaia-Enceladus was truncated $9.6 \pm 0.2$~Gyr ago. This moment marks the final stage of the Gaia-Enceladus merger. Values for this age between 8-10 Gyr have previously been found in the literature \citep{gallart2019NatAs...3..932G,montalban2021NatAs...5..640M,Grunblatt2021,Borre22}, but here this moment is dated with greater precision. No star in Gaia-Enceladus was found to continue the sequence to metallicities higher than about $-0.7$~dex; this is similar to the findings of \cite{feuillet2021MNRAS.508.1489F}. Simulations have shown that after a gas-rich merger, a thin disk can form from gas that has the time to be polluted by Type Ia supernovae \citep{Brook2007}. Although we cannot confirm the causal connection here, our dating for the Gaia-Enceladus merger is consistent with the idea that its gas will be used to form the low-$\alpha$ thin disc population \citep{Bonaca20}, since these stars mostly have younger ages \citep{haywood2018ApJ...863..113H}.

In Fig.~\ref{fig:feh.age}, we investigate the time evolution of the chemical enrichment in other metal-poor stellar populations identified in our sample. The plots focus on the so-called \spl\ \citep{belokurov2020MNRAS.494.3880B} and on a population that we call \Ere. The \spl\ is made of old, somewhat metal-rich stars (mainly with [Fe/H] $>-0.8$), probably formed in the thick disc of the Milky Way. Its stars mostly have prograde motions and were thrown into high-eccentricity orbits during the accretion of Gaia-Enceladus \citep{Bonaca2017,diMatteo2019A&A...632A...4D,belokurov2020MNRAS.494.3880B}. \Ere, on the other hand, is a population of more metal-poor stars (mostly [Fe/H] $<$ $-$0.8), mostly in slightly retrograde orbits, although also including a wide tail of stars with prograde orbits (see Fig.~\ref{fig:Splash_thamnos}). 

As seen in Fig.~\ref{fig:feh.age}, the age-metallicity sequence of the \spl\ stars overlaps that of thick-disc stars. Our precise age dating clearly shows that most of the \spl\ stars formed before the final stages of the Gaia-Enceladus merger. The peak of their age distribution is around 11 Gyr, and the mean metallicity is around [Fe/H] $\sim$ $-$0.7. After the peak, the age distribution shows a decline for about 2 Gyr, with its minimum roughly coinciding with the age truncation of Gaia-Enceladus. Subsequently, a slight increase is observed. However, the significance of this increase is unclear. \spl\ stars were only found in the GALAH sample, for which we obtained ages with slightly higher mean uncertainty.

Regarding \Ere, and as several halo substructures have already been identified and named in the literature \citep[e.g.,][]{koppelman2019A&A...631L...9K,Myeong2019,Naidu2020,Horta21,Lovdal22}, a valid question is whether we have rediscovered a substructure that is already known. In the region occupied by \Ere\ in the mid-panel of Fig.\ \ref{fig:Splash_thamnos}, there are two main substructures identified in the literature, Sequoia \citep{Myeong2019} and Thamnos \citep{koppelman2019A&A...631L...9K}. The stars of Sequoia are usually thought to occupy a region of loosely bound stars with energy $>$ $-$1.6 $\times$ 10$^5$ km s$^{-2}$ in the Lindblad diagram. In this region, we only have a handful of objects. Thamnos stars, as defined by \citet{koppelman2019A&A...631L...9K}, would be made of two components with, among other properties, $V_{\phi}$ centred at $-$150 and $-$200 km s$^{-1}$. Instead, what we call \Ere\  appears strongly concentrated at $V_{\phi}$ $>$ $-$100 km s$^{-1}$ (see Fig. \ref{fig:populations}), although we deduce by the analysis in Fig.~\ref{fig:Splash_thamnos} that part of it remains masked by the \spl\ rotation distribution, composing a dynamic structure with virtually null net rotation $L_Z \approx 0$. We cannot exclude the possibility that a few Sequoia and/or Thamnos stars are part of our \Ere\ selection, but it seems clear that \Ere\ is not dominated by stars belonging to these substructures.

An important point to note here is that our sample is very small (228 stars with precise ages) compared to large samples, with thousands or tens of thousands of stars, used by works that search for halo substructures. Therefore, we are not in a position to identify overdensities of stars that distinguish themselves from a general background stellar population. In contrast, what we call \Ere\ is, most likely, exactly such background stellar population. As we discuss below, the age-[Fe/H]-[$\alpha$/Fe] relations of these stars suggest that they are most likely of in situ origin and a significant fraction of them acquired retrograde orbits after being subject to the effects of several mergers in the history of the Milky Way \citep{Jean-Baptiste2017}.

Figure~\ref{fig:feh.age} shows that the stars of \Ere\ attained a higher metallicity earlier than the stars from Gaia-Enceladus. This fact suggests that the \Ere\ stars formed in a system or region that could sustain star formation with higher efficiency. Moreover, \Ere\ stars consistently show a high [Mg/Fe] up to [Fe/H] $\sim$ $-$0.5. Only around this metallicity does the SNIa contribution become apparent. Together, these facts suggest that \Ere\ originated in a system that was more massive than Gaia-Enceladus. The most obvious candidate is the Milky Way itself, where the knee in [Mg/Fe] ratios also appears at [Fe/H] $\sim$ $-$0.5. 

The age distribution of \Ere\ stars peaks at a value of $\sim$11.5 Gyr, slightly older but still similar to the \spl. Interestingly, essentially no star in this group was found to be younger than the age of the final stage of the Gaia-Enceladus merger. Because the \Ere\ stars are old and mostly retrograde, it is possible that their orbits show the combined effect of heating caused by several mergers. As discussed in \cite{Jean-Baptiste2017}, the higher the number of mergers suffered by the host galaxy, the broader the area in the retrograde region of the Lindblad diagram (see e.g.~Fig.~\ref{fig:Splash_thamnos}) that the in situ stars will end up occupying. Alternatively, there are at least two other possibilities for the origin of the \Ere\ stars: i) the core region of Gaia-Enceladus itself and ii) a separated merger with a galaxy more massive than Gaia-Enceladus.

The simulations of \cite{Koppelman2020} demonstrated that debris from a merger with a disc galaxy \citep[as was probably the Gaia-Enceladus progenitor,][]{helmi2018} have a final complex phase-space structure, a range of orbital properties, and also stars with a range of chemical abundances. In the core regions of disc galaxies, the star formation is more intense. Consequently, as seen in the Milky Way, stars formed in the inner regions can be old, metal rich, and have enhanced [Mg/Fe] \citep{Trevisan2011}. In that sense, what we call \Ere\ could be a population of stars from the core of Gaia-Enceladus. The decreasing age distribution of \Ere, after 11.5 Gyr, and potential cut-off around 9.5 Gyr, could be understood as being influenced by the then on-going merger. However, the simulations of \cite{Koppelman2020} also show that for the assumed configuration of the Gaia-Enceladus merger, stars from the core region would end up with prograde orbits of low eccentricity, which is not the case for the \Ere\ stars. Thus, it seems that this hypothesis for the origin of \Ere\ can be excluded.

Another possibility is that of a separated merger with a system that was more massive than the Gaia-Enceladus galaxy. However, as we discussed above, the kinematic and orbital properties of the Erebus stars do not seem to match the substructures identified in the same region of the Linblad diagram in other works \citep{koppelman2019A&A...631L...9K,Naidu2020,Shank2022}. On the other hand, one should also keep in mind that a single merger can give origin to several substructures \citep{Jean-Baptiste2017}. To investigate any connection between \Ere\ and other halo substructures, we will need to extend our precise analysis of abundances and ages to stars from these other substructures for comparison. What we do observe is that the age-metallicity relation for \Ere\ stars is distinct from that of Gaia-Enceladus. At this point, a connection between the two populations seems unlikely.

Here, we highlight a possible connection with the simulations of galaxy formation in a cosmological context considered in \cite{belokurov2020MNRAS.494.3880B}. As those authors discussed, some of the Auriga simulations where a host galaxy had a proto-disc produced a stellar population with properties consistent with those of the \spl\ stars after a merger (where the heated stars managed to retain much of their angular momentum). However, in one of the simulations, the signature of a prograde \spl\ was less clear. The authors speculated that the original configuration of the heated stars was less of a disc and more spheroidal. This could be a possible explanation for the \Ere\ stars. This population could be a component that was already in a spheroid configuration but that suffered additional heating by the Gaia-Enceladus accretion and perhaps also by other previous merger events. This possibility lends additional support to our conclusion that \Ere\ is a stellar population of in-situ origin. Because of the complex nature of this population and its somewhat "chaotic" orbital characteristics, we got inspired to name it \Ere, a primordial deity in Greek mythology born out of Chaos.

A similar in-situ stellar component dubbed \textit{Aurora} has been recently identified by \citet{2022MNRAS.514..689B}. The Aurora was defined with a metallicity cut of [Fe/H] $>$ $-$1.5 and a strict selection of energy values (on a scale quite different from the one used here, making a comparison difficult). Aurora stars were found to have a modest net rotation. This is different from our selection, which includes more metal-poor stars without restriction in energy, although it also appears to be of null net rotation. We cannot exclude that these are two sides of the same in-situ population, with \Ere\ being a more general and extended population that encompasses Aurora. We also note that a suggestion has been made by \citet{2022ApJ...938...21M} that Aurora is connected to the Heracles structure identified by \citet{Horta21} in the inner 4 kpc of the Galaxy (which is a region not probed by our stellar sample). On the other hand, \citet{Horta22} finds clear chemical differences between stars associated with Aurora and Heracles. As a connection between \Ere\ and all these populations is, at least for now, not clear, we believe that giving it a separated identification is justified. Precise ages for stars in all these groups will be needed to clarify whether their age-chemical relations are distinct or not.

Finally, we mention that the plots in Fig.~\ref{fig:feh.age} include stars with orbits and [$\alpha$/Fe] ratios consistent with those in the thin disc but with atypically old ages \citep{haywood2013A&A...560A.109H,haywood2019A&A...625A.105H}. Judging by their values of age, metallicity, and [$\alpha$/Fe], these stars seem to be part of a coeval population of high prominence in the inner disc (up to 10~kpc radius) identified by \cite{Ciuca2022}.
On the basis of comparisons with the cosmological simulations from \citet{Grand2020}, \cite{Ciuca2022} interpret this population as evidence of a starburst generated by gas accretion induced by the merging with Gaia-Enceladus. This new generation of stars then has ages roughly consistent with those of the \spl\ and a range of [Fe/H] values that are lower than typical values of thin-disc stars due to dilution with the accreted metal-poor gas.

The discussion above shows how much the precise ages and abundances derived in this work help to put together all the pieces of the old Milky Way populations in a consistent chronological order. We can clearly detect the halo in-situ stellar population that existed before the Gaia-Enceladus merger and was heated to more retrograde orbits. We can also detect the stellar population of the disc that was affected by the same merger. Finally, we can also precisely date the final stages of the Gaia-Enceladus merger, at 9.6 $\pm$ 0.2 Gyr. This dating is also consistent with the idea that the gas from Gaia-Enceladus helps to form the low-$\alpha$ population of the thin disc \citep{Brook2007,Bonaca20}. The data produced in this paper used for the discussion in this section are provided as supplementary material at CDS.

\begin{acknowledgements}
    REG and RS acknowledge the support of the National Science Centre, Poland, through project 2018/31/B/ST9/01469. REG also acknowledges the support by Fonds de la Recherche Scientifique (F.R.S.-FNRS) and the Fonds Wetenschappelijk Onderzoek-Vlaanderen (FWO) under the EOS Project nr O022818F. REG also acknowledges F\'elix Isidoro Escate Barrios and Margarita Santamaria Vel\'asquez for being sources of inspiration since the conception of this work; they will be kept in memory although their prompt departure. REG acknowledges Andr\'e Rodrigo da Silva and Maria Luiza Dantas for fruitful discussions of the paper's topic. Use was made of the Simbad database, operated at the CDS, Strasbourg, France, and of NASA’s Astrophysics Data System Bibliographic Services. 
    This publication makes use of data products from the Two Micron
    All Sky Survey, which is a joint project of the University of
    Massachusetts and the Infrared Processing and Analysis
    Center/California Institute of Technology, funded by the National Aeronautics and Space Administration and the National Science Foundation.
    This research used Astropy,\footnote{http://www.astropy.org} a community-developed core Python package for Astronomy \citep{astropy:2018}.
    This work presents results from the European Space Agency (ESA)
    space mission Gaia. Gaia data are processed by the Gaia Data Processing and Analysis Consortium (DPAC). Funding for the DPAC is provided by national institutions, in particular the institutions participating in the Gaia MultiLateral Agreement (MLA). The Gaia mission website is \url{https://www.cosmos.esa.int/gaia}. The Gaia archive website is \url{https://archives.esac.esa.int/gaia}.
\end{acknowledgements}

\bibliographystyle{aa.bst}

\bibliography{Faint2}

\begin{appendix} 
\onecolumn
\section{Complementary figures}
\label{app:abundance_calibration}

\begin{figure*}
    \centering
    \includegraphics[width=0.8\linewidth]{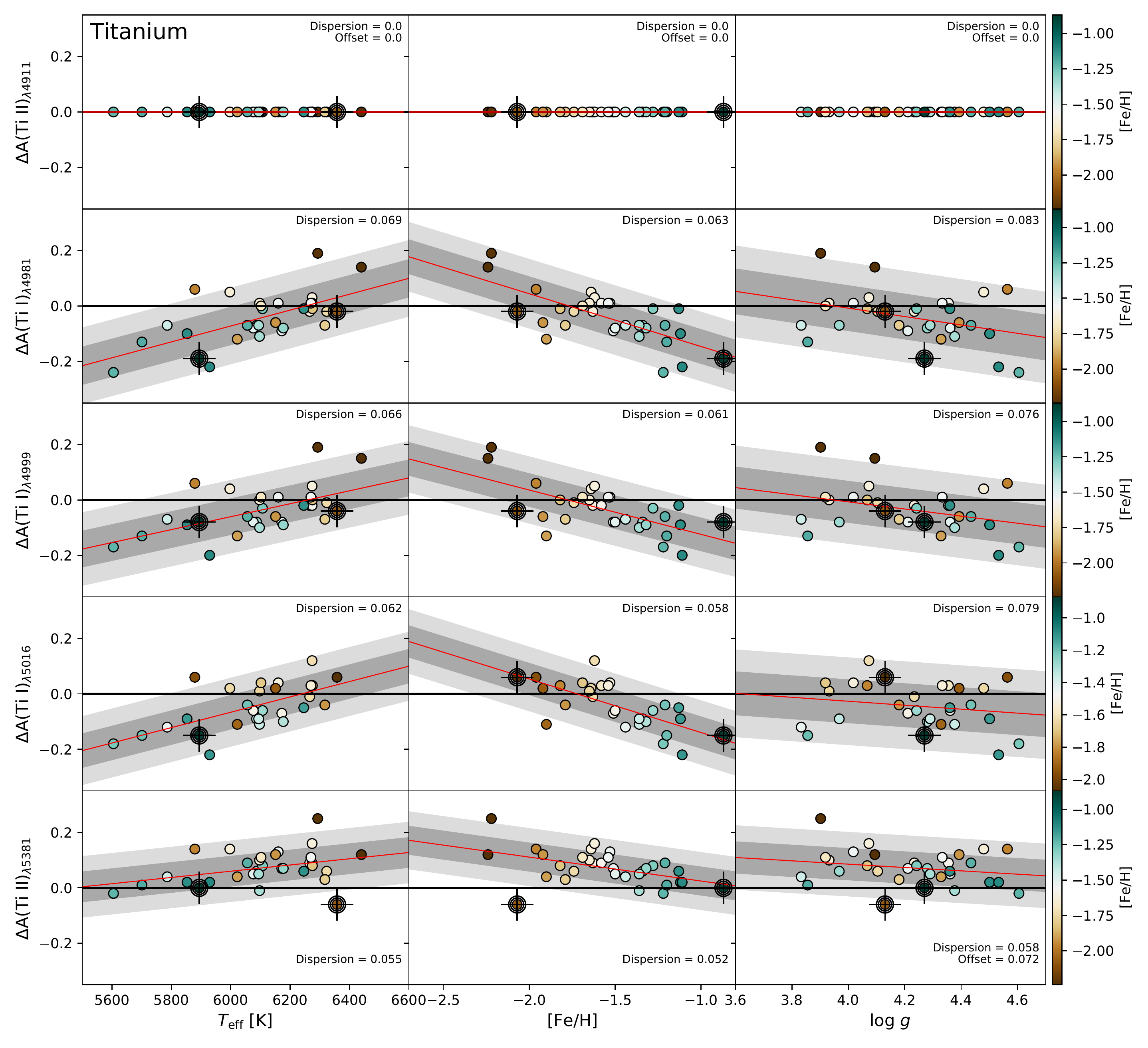}
    \caption{Similar to Fig.~\ref{fig:Mg_comparison} for titanium. Trends are computed with abundances relative to those of the line 4911~\AA.}
    \label{fig:Ti_comparison}
\end{figure*}

\begin{figure*}
    \centering
    \includegraphics[width=0.8\linewidth]{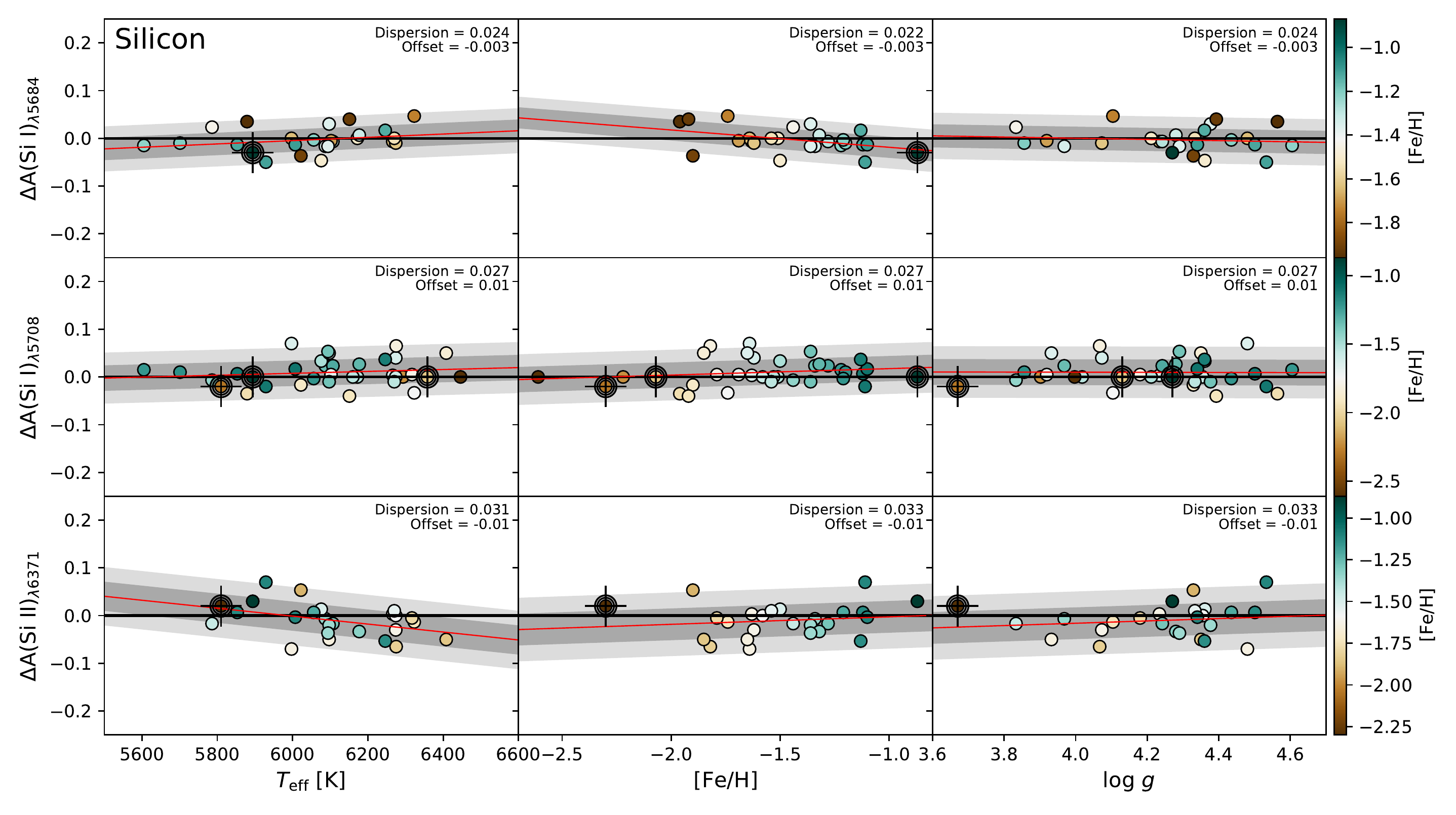}
    \caption{Similar to Fig.~\ref{fig:Mg_comparison} for silicon. Trends are computed with abundances relative to average values.}
    \label{fig:Si_comparison}
\end{figure*}

\begin{figure*}
    \centering
    \includegraphics[width=0.8\linewidth]{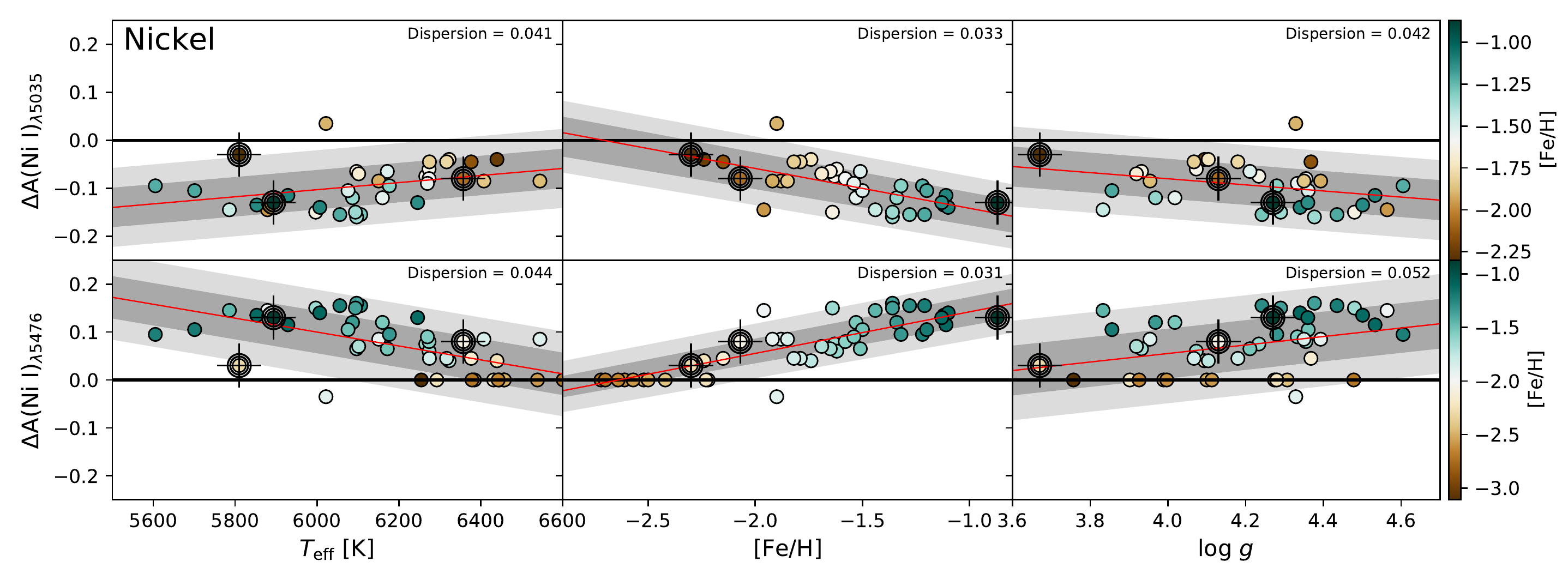}
    \caption{Similar to Fig.~\ref{fig:Mg_comparison} for nickel. For each line, the abundances plotted are relative to those of the other line.
    For the line 5476~\AA, relative abundances that do not have measurements available from the line~5035~\AA\ are set to zero.}
    \label{fig:Ni_comparison}
\end{figure*}

\begin{figure*}
    \centering
    \includegraphics[width=0.8\linewidth]{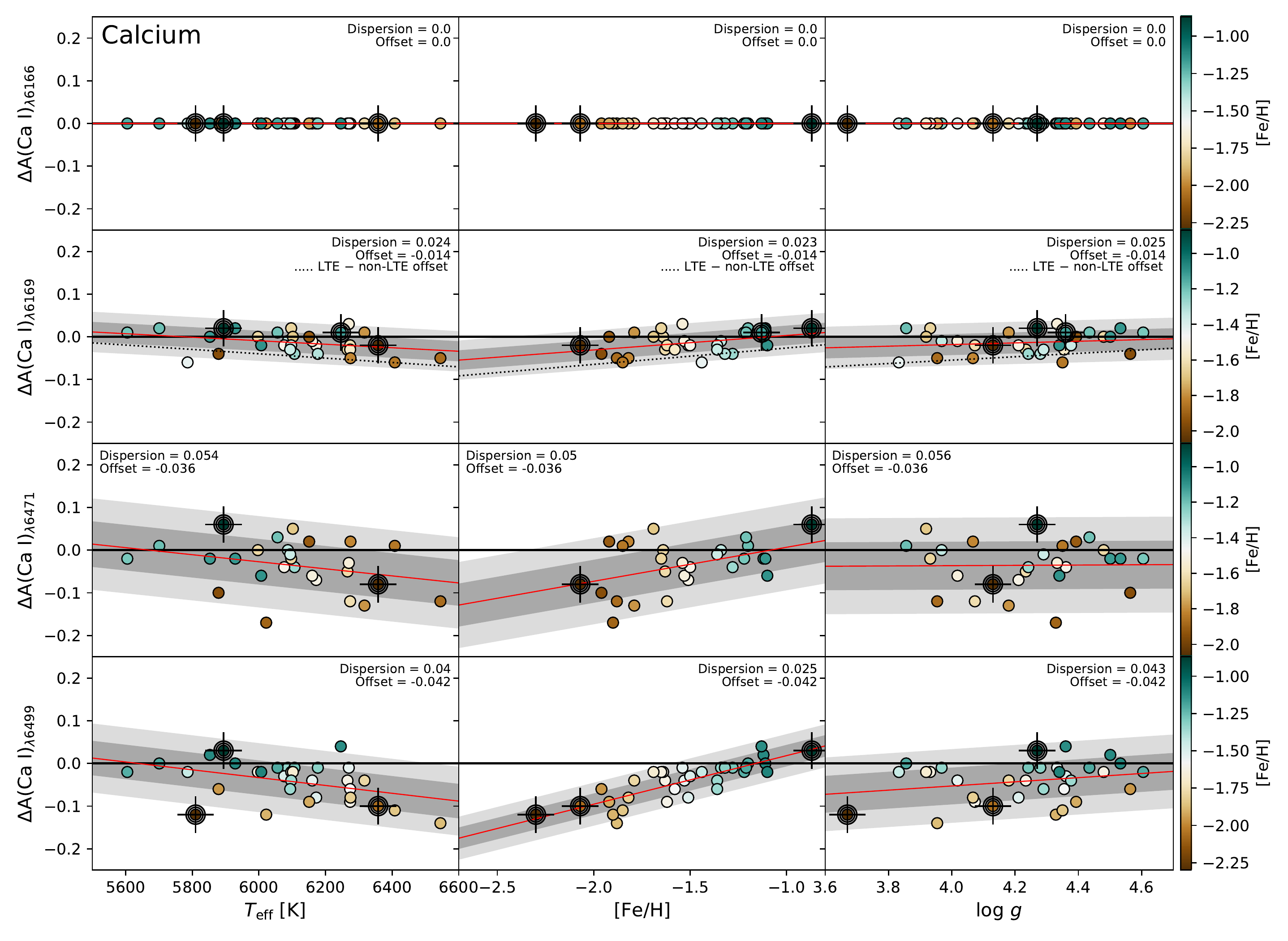}
    \caption{Similar to Fig.~\ref{fig:Mg_comparison} for calcium.
    Trends are computed with abundances relative to those of the line 6166~\AA.}
    \label{fig:Ca_comparison}
\end{figure*}

\begin{figure*}
    \centering
    \includegraphics[width=0.8\linewidth]{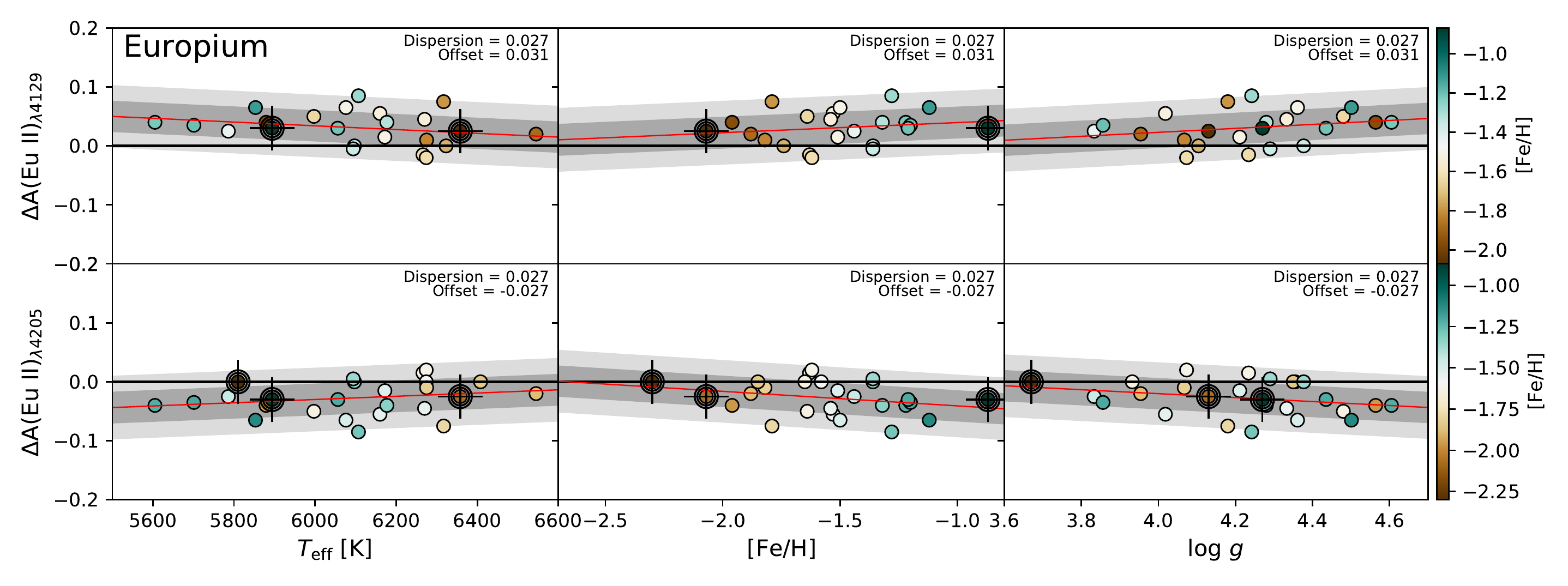}
    \caption{Similar to Fig.~\ref{fig:Ni_comparison} for europium.
    For each line, the abundances plotted are relative to those of the other line. For the line 4205~\AA, relative abundances that do not have measurements available from the line~4129~\AA\ are set to zero.}
    \label{fig:Eu_comparison}
\end{figure*}

\begin{figure*}
    \centering
    \includegraphics[width=0.8\linewidth]{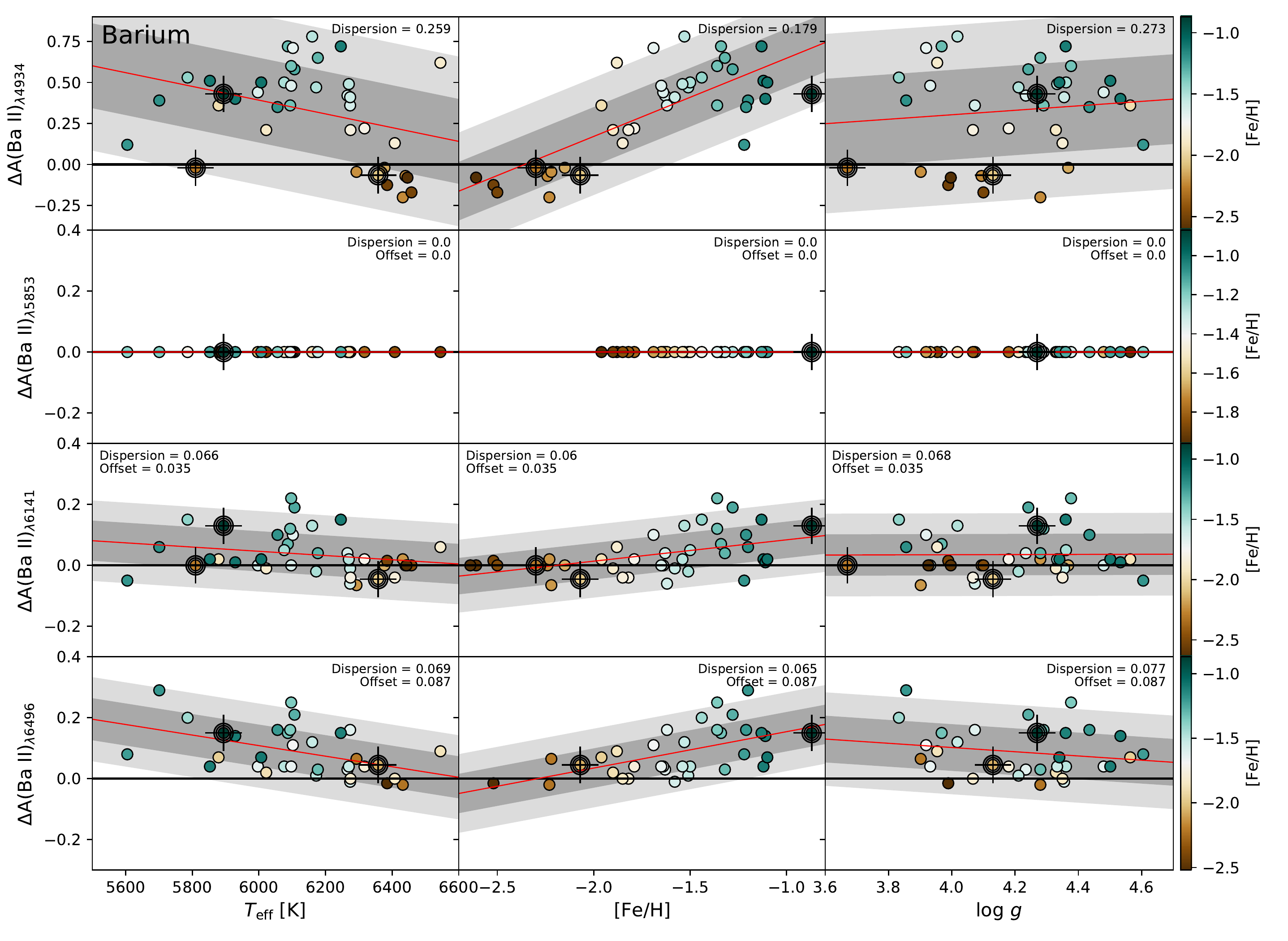}
    \caption{Similar to Fig.~\ref{fig:Mg_comparison} for barium. Trends are computed with abundances relative to those of the line 5853~\AA.}
    \label{fig:Ba_comparison}
\end{figure*}

\begin{figure*}
    \centering
    \includegraphics[width=0.3\linewidth]{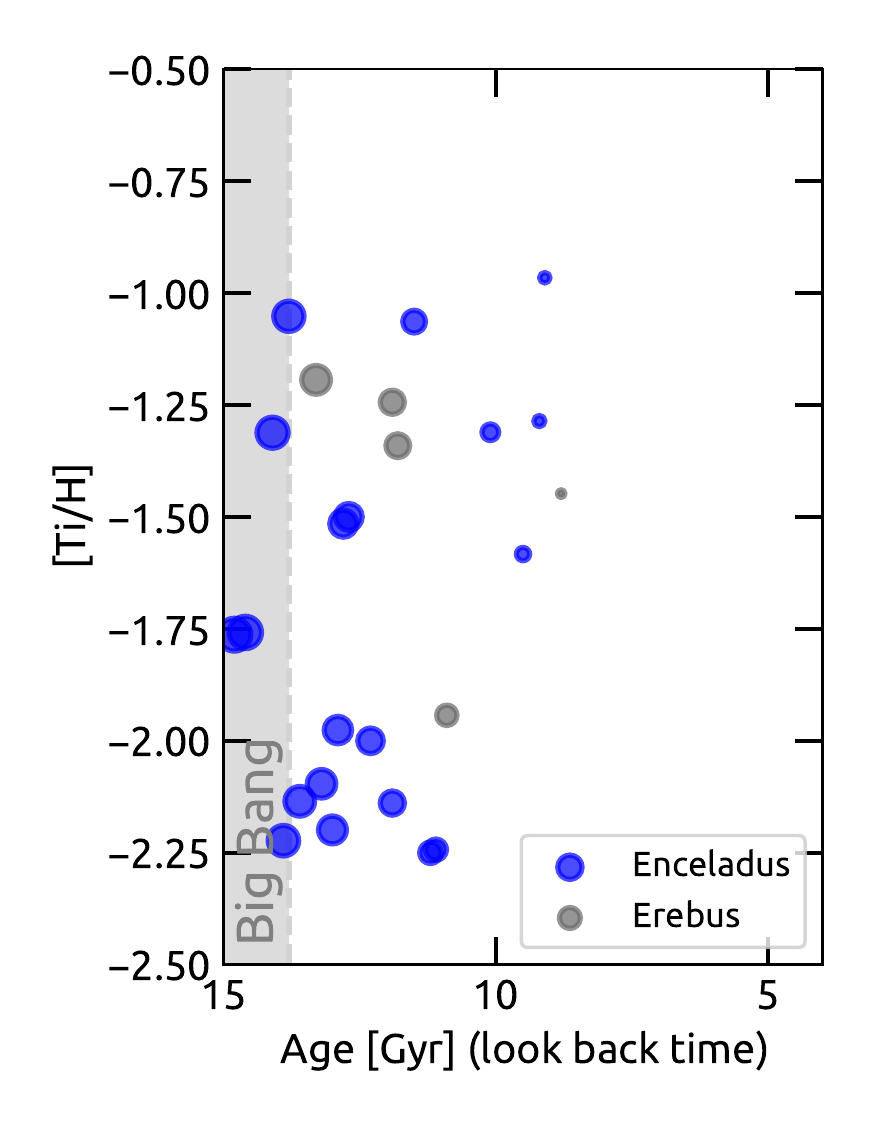}
    \includegraphics[width=0.3\linewidth]{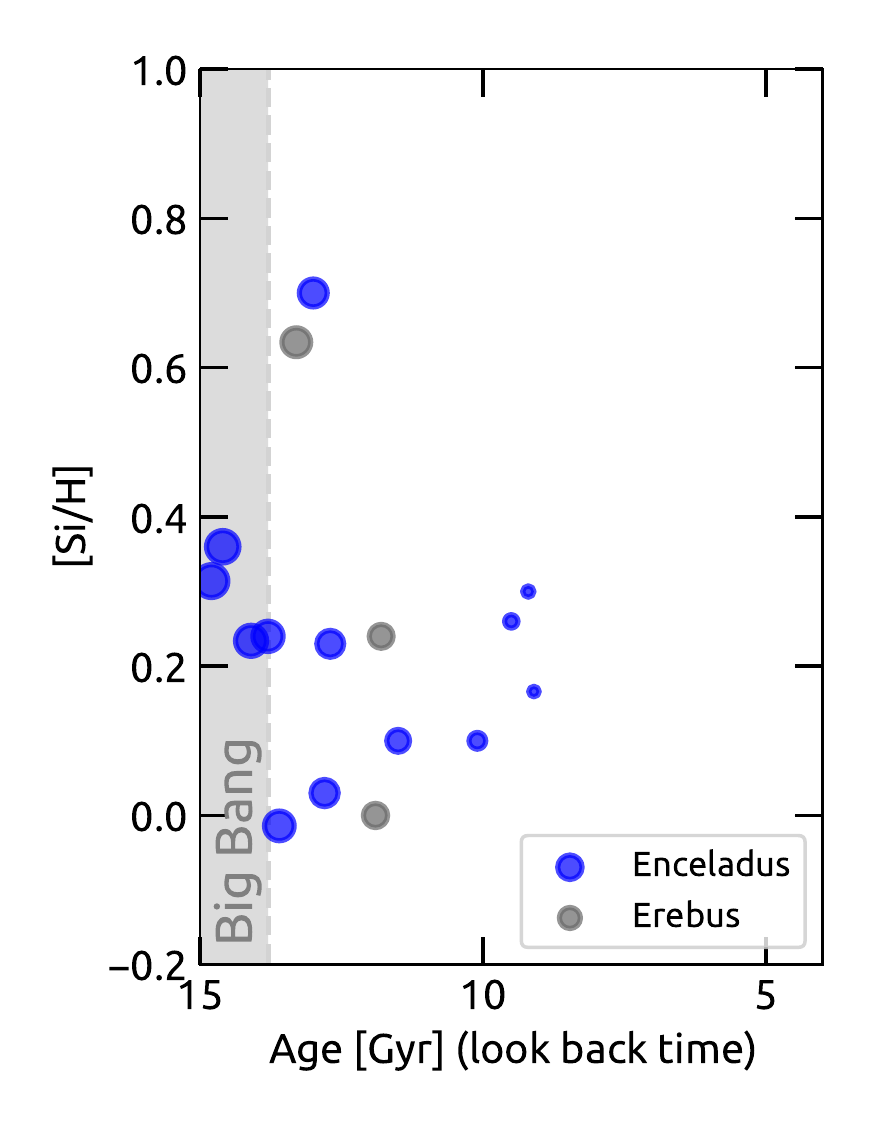}
    \includegraphics[width=0.3\linewidth]{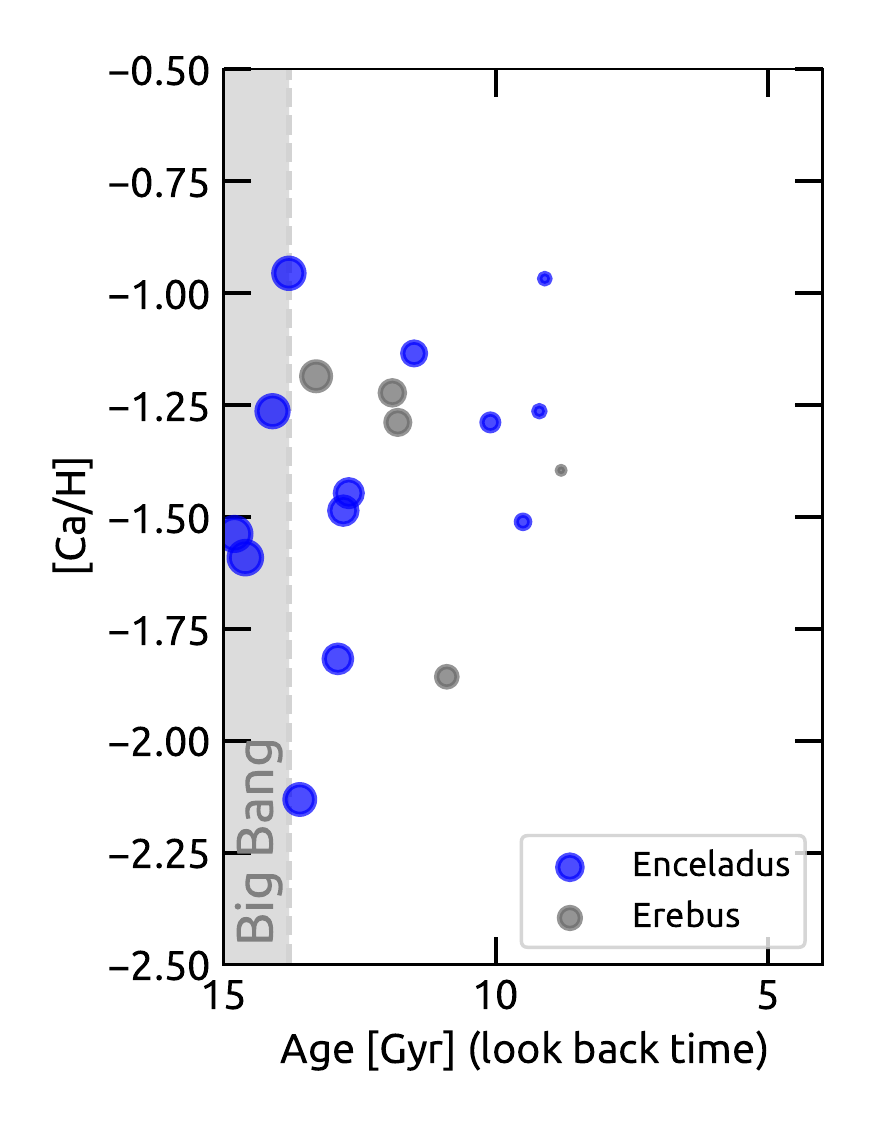}
    \includegraphics[width=0.3\linewidth]{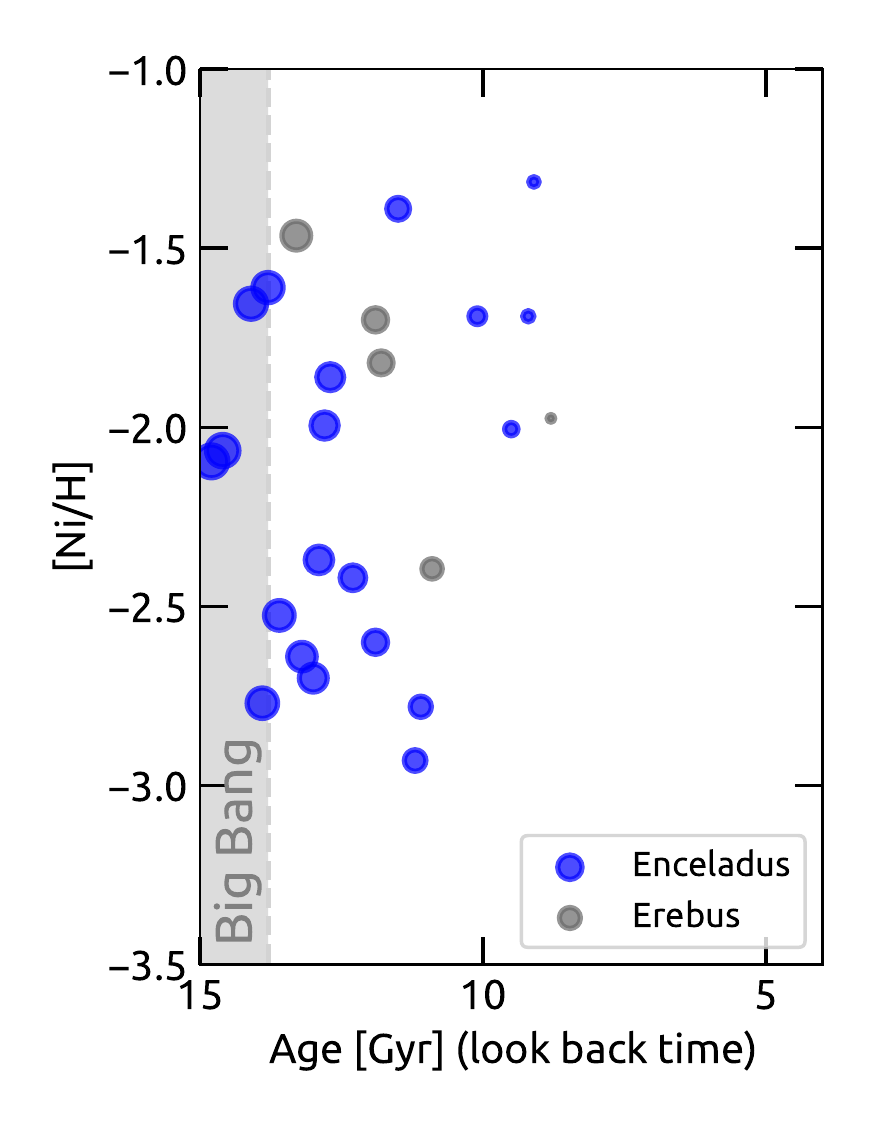}
    \includegraphics[width=0.3\linewidth]{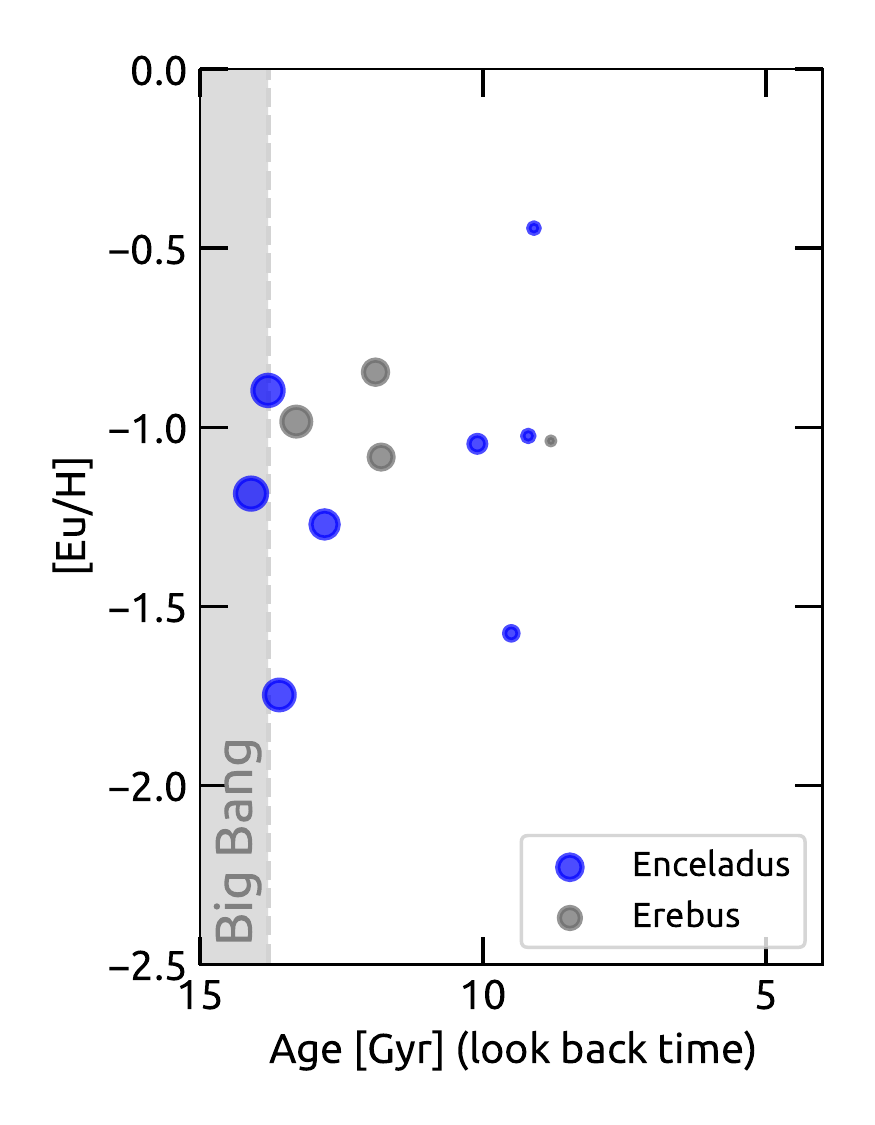}
    \includegraphics[width=0.3\linewidth]{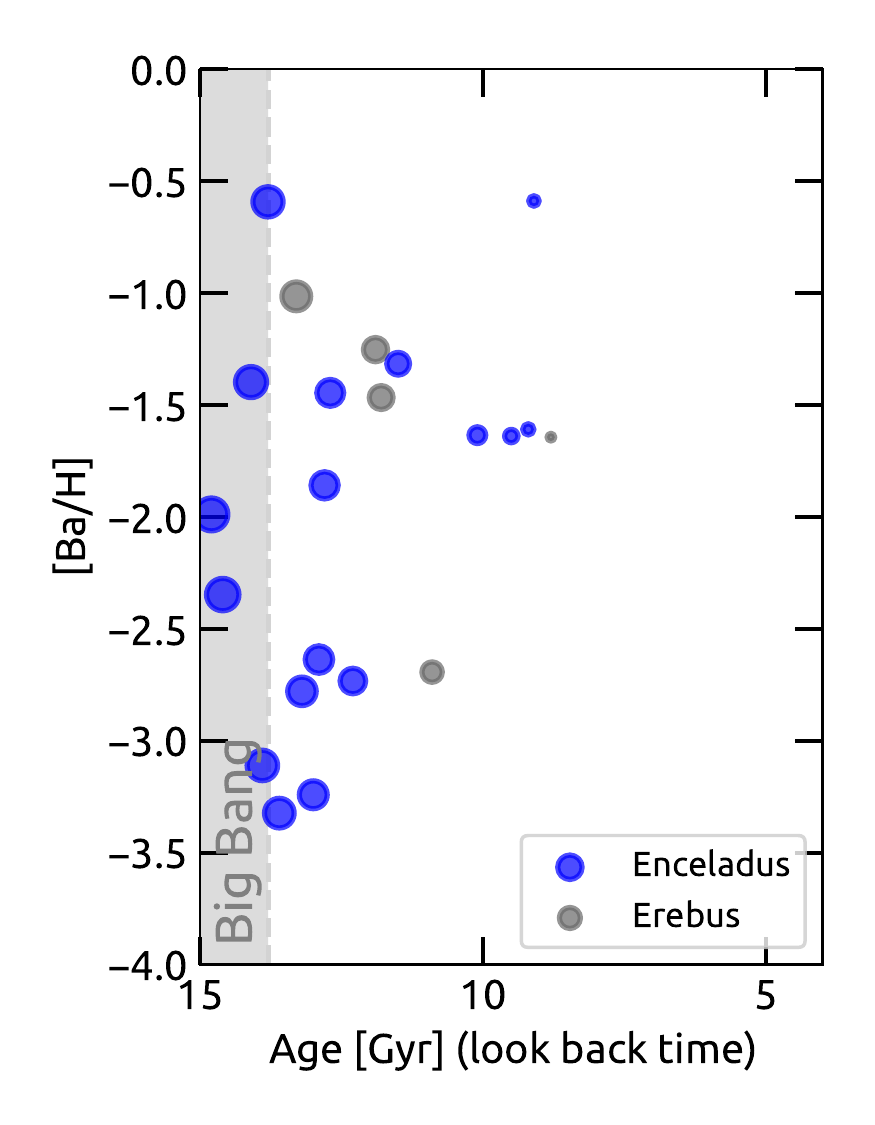}
    \caption{Element-hydrogen ratios as function of the stellar age for the \titan\ with halo dynamics. The symbol size changes as a function of stellar age.}
    \label{fig:other_elements}
\end{figure*}

\end{appendix}

\end{document}